\documentclass[letterpaper,twocolumn,10pt]{article}

\usepackage{usenix-2020-09}
\usepackage{cite}
\usepackage{amsmath,amssymb,amsfonts}
\usepackage{algorithmic}
\usepackage{graphicx}
\usepackage{textcomp}
\usepackage{makecell}
\usepackage{subcaption}
\usepackage{balance}
\usepackage{enumitem}
\usepackage{wasysym}
\usepackage{pifont}
\usepackage{multirow} 
\usepackage{tabularx} 
\usepackage{booktabs}
\usepackage[affil-it]{authblk}
\newcommand{\cmark}{\ding{51}}%
\newcommand{\xmark}{\ding{55}}%

\begin{document}

\title{Unveiling Dynamic Binary Instrumentation Techniques}

\author[1,2]{Oscar Llorente-Vazquez}
\author[3]{Xabier Ugarte-Pedrero}
\author[4]{Igor Santos-Grueiro}
\author[2]{Pablo Garc\'ia Bringas}
\affil[1]{TECNALIA, Basque Research and Technology Alliance (BRTA)}
\affil[2]{Faculty of Engineering, University of Deusto}
\affil[3]{Cisco Systems, Inc.}
\affil[4]{HP Labs}

\maketitle

\begin{abstract}
Dynamic Binary Instrumentation (DBI) is the set of techniques that enable
instrumentation of programs at run-time, making it possible to monitor and
modify the execution of compiled binaries or entire systems.
DBI is used for countless security applications and analyses, and is
extensively used across many fields in both industry and academia.
Over the years, several DBI approaches have been proposed based on
different technologies and implementing diverse techniques. Every solution tries
to overcome certain limitations, but they bring other
shortcomings. Some are specialized for one particular domain or task, while
others have a wider scope.

In this paper, we shed light into the labyrinth of DBI, bringing together
process-level and whole-system approaches. We depict their building blocks
and analyze the underlying instrumentation techniques, comparing their
ability to instrument different primitives and run-time events. Then, we
evaluate their performance over a comparative framework when implementing each primitive,
and highlight relevant observations. Our results show that no single
technique is better than the rest in all circumstances.
\end{abstract}

\section{Introduction}
\label{sec:introduction}

Dynamic Binary Instrumentation (DBI) frameworks are an extremely versatile and
powerful mechanism with a wide range of applications. For instance, they have
been extensively used for performance profiling and
analysis~\cite{ketterlin2012profiling,attariyan2012x} or software fault
injection~\cite{wei2014quantifying}. Furthermore, prior work~\cite{d2019sok} has
highlighted the pervasiveness of DBI in security research, which includes
applications such as
shepherding to enforce security policies such as Control-Flow Integrity
(CFI)~\cite{prakash2015vfguard,van2016tough}, vulnerability discovery
and analysis~\cite{jurczyk2013identifying,xiao2017stacco}, fuzzing of embedded
devices~\cite{pustogarov2020ex,clements2020halucinator}, and malware
analysis~\cite{polino2017measuring,ugarte2016rambo}.

The main goal of DBI is to enable the execution of analysis callbacks that can
inspect and tamper with the execution at arbitrary points in a running program.
However, it is not clearly established what DBI comprises in terms of its core
elements, the execution contexts and the scope. This often leads to only
consider approaches that insert and execute code as part of the target program
under analysis from a user-space perspective.

Due to the high number of DBI implementations, and the heterogeneity in the
designs and techniques, the current state of the art is diffuse, hard to
understand and properly evaluate. In addition, many DBI tools have steep
learning curves and require deep expertise to understand their internals,
configure and use them effectively. The DBI landscape is fragmented across
several research papers, websites and source code repositories, and
documentation quality varies. It also lacks benchmarks for DBI tools across
common metrics. This leads to additional challenges in order to understand the
fundamentals and capabilities of existing solutions, to know which techniques
and approaches are adequate in which circumstances, and to understand the
advantages and disadvantages of each of them.

On the one hand, there has been extensive discussion on DBI transparency
and evasive
behavior~\cite{miramirkhani2017spotless,d2019sok,filho2022evasion,d2022evaluating},
especially in the context of adversarial environments such as malware analysis,
but to the best of our knowledge there is no existing work that clearly
categorizes DBI components, and analyzes and compares existing instrumentation
techniques, their capabilities and relative performance in a comprehensive
manner. In fact, they usually focus on a single technique or technology, or
exclusively consider process-level or whole-system approaches.
For instance, in a previous work~\cite{d2019sok}, authors focus on DBI
evasion and escape problems in the context of Just-In-Time (JIT) dynamic
compilation based user-space DBI. Our goal is different in nature -- instead of
focusing on the transparency and security of specific solutions, we seek to
analyze and evaluate all known instrumentation techniques and their
capabilities as the core of DBI.

On the other hand, even though performance overhead is one of the main concerns
and challenges in DBI, a fair amount of research papers that propose a new
technique or DBI tool do not evaluate its performance, measure different
metrics, or only report the relative overhead with respect to native execution
using arbitrary test cases. Unfortunately, this is not sufficient to understand
which approach might show better performance for a particular situation. In
contrast, we take into consideration the existing techniques, and analyze their
performance when instrumenting different types of events under different
conditions, under a variety of programs, with different lengths, and a different
frequency of observable events.

In this paper, we fill this gap with a comprehensive and systematic study of
DBI. Specifically, this work dissects the building blocks of DBI, and
identifies, classifies, analyzes, and evaluates the existing instrumentation
techniques, pointing out their different characteristics and capabilities. We
evaluate the different techniques by testing a set of reference
implementations over a complete comparative framework. Our analysis does not
pretend to compare tools or implementations against each other, since
their capabilities and performance are often influenced by design and
implementation decisions (e.g., inserting instrumentation
routines written in languages such as Python or Javascript, which are
significantly less efficient than C/C++). Instead, we focus on the underlying
instrumentation techniques upon which existing DBI frameworks are built.

In particular, the contributions of this work are:
\begin{itemize}
    \item The definition and drilldown of the DBI building blocks. Namely,
        instrumentation, introspection and execution manipulation.

    \item A comprehensive characterization and classification of existing
        instrumentation techniques in relation to the fundamental
        instrumentation primitives they can provide. We further compare existing
        DBI frameworks that implement each technique, covering process-level and
        whole-system approaches.

    \item A thorough performance evaluation upon our framework of the
        instrumentation techniques for each instrumentation primitive, over a
        wide range of programs and conditions. Our findings show that the
        performance of the techniques often depend on factors such as
        the nature of the program and the frequency of the triggered
        instrumentation events during execution, and that no single technique
        stands out from the rest in every possible test.
\end{itemize}

The remainder of this paper is organized as follows.
\textsection\ref{sec:bblocks} depicts the underlying principles of DBI, namely,
instrumentation, introspection, and manipulation.
\textsection\ref{sec:tech} classifies and details the inner workings of the
different instrumentation methods. \textsection\ref{sec:map} describes the
existing tools that implement those techniques and the variety of run-time
events they can instrument.
Then, \textsection\ref{sec:eval} evaluates the performance of the
instrumentation techniques in different scenarios by means of their practical
implementations.
\textsection\ref{sec:insights} interprets and reflects on the implications of the
results obtained from the evaluation.
\textsection\ref{sec:dis} discusses the main contributions of this work and
reviews their impact, while \textsection\ref{sec:relatedwork} reviews the
different works that are related to our paper. Finally,
\textsection\ref{sec:conc} concludes the paper.

\section{DBI Building Blocks}
\label{sec:bblocks}

We define DBI systems as those that take control of the execution of a target
program, allowing to execute user-defined code on different execution events
or with different granularity, making it possible to examine the program
state and alter the execution of the program in an automated way.

Considering the proposed definition, we may argue that every DBI system has 3
fundamental building blocks that also determine their capabilities: (i) one or
more instrumentation techniques that enable execution control, including event
generation and callback registration mechanisms, (ii) the introspection
capability (i.e., the ability to observe the architectural state of the target),
and (iii) the capability to alter the execution of the target, either directly
or by modifying the program state.

\subsection{Process-Level \& Whole-System Approach}

Depending on the scope, DBI systems are classified into process-level and
whole-system. Process-level DBI includes every approach whose scope is
restricted to a single process usually in user-space, being only able to manage
one program at a time.  The DBI runtime can reside in the same address space as
the target or in a remote location, so they can spawn or attach to a running
process from the
user-space~\cite{bruening2004rio,luk2005pin,nick2007valgrind,bernat2011anywhere,frida,gdb}
or kernel-space~\cite{vasudevan2006cobra}. On the contrary,
whole-system solutions recreate entire systems and are 
composed of a host and a guest system. Therefore, instrumentation and analysis
logic is generally implemented out of the analyzed
target~\cite{yan2012v2e,lengyel2014drakvuf,dolan2015panda,chipounov2011s2e,zhang2015malt}.

Each approach comes with its own benefits and disadvantages. Whole-system tools
provide a full-system view, allowing to observe kernel activity and process
interaction. However, there is a clear increased performance cost compared to
process-level approaches. Process-level DBI often share space with the
instrumented program, while whole-system tools perform the instrumentation and
analysis from outside the virtual machine, providing better transparency and
stronger isolation. However, since a new layer is introduced between the host
and the guest system, a significant challenge emerges regarding introspection
--- the semantic gap problem~\cite{jain2014sok}.

\subsection{Instrumentation}
The backbone of DBI is formed by the different instrumentation methods that
enable the controlled execution of a target program and the execution of
user-defined code.

Using several techniques such as inserting breakpoints, trampolines, or 
dynamic compilation, DBI systems can interleave code or temporarily stop the
execution at specific points (e.g., when a breakpoint is hit, or whenever an
instruction sequence is compiled), or when certain events are triggered (e.g.,
interactions with the OS). These systems offer a set of instrumentation
primitives (i.e., facilities through an API that allow the user to run callback
routines at different granularities). Depending on the technique and, in
certain cases, the Instruction Set Architecture (ISA) abstractions provided by
the DBI framework, diverse events are exposed, such as the execution of
instructions, memory accesses, or OS and hardware events.
Moreover, the design of the DBI engine determines whether these user-defined
callbacks are executed in the same context as the target under analysis,
in a separate process (i.e., a debugger process), or in a higher privilege
context like the kernel (Ring 0) or the hypervisor (Ring -1).

Instrumentation techniques, capabilities, and related primitives are classified
and further detailed in section \textsection\ref{sec:tech}.

\subsection{Introspection}
Another relevant aspect of DBI is related to the ability to observe the program
state at different points of the execution.

In general, DBI frameworks provide means through the API to access the CPU
register state and memory contents. Similarly, they may also facilitate
introspection by means of higher level abstractions passed to analysis calls as
arguments. Besides, some tools may provide complementary methods to
access devices like hard disk drives, and so on.
Even though register values and raw memory contents are sufficient for some
tasks, a high-level representation closer to the original source, such as
variables, data structures, or functions, is extremely useful in many cases
(e.g., vulnerability analysis, data manipulation). To this end, the conventional
method is to leverage standard debugger techniques that explore the symbol
table and debugging information if present~\cite{gdb,idapro}.

In whole-system approaches, introspection poses a greater challenge
since they must confront the semantic gap problem. The semantic gap is defined
as a disparity in abstractions. It fundamentally arises when the DBI system has
an external point of view (e.g., at hypervisor-level), and high-level OS
abstractions, such as threads and processes are not visible. In contrast,
hardware-level abstractions such as physical memory pages and hardware device
operations can be observed. To bridge the semantic gap, numerous research works
have been proposed in the area of Virtual Machine Introspection
(VMI)~\cite{dolan2011virtuoso,fu2012space,saberi2014hybrid,zhao2017seeing}. This
process tries to retrieve information about the current state of the guest by
reconstructing high-level semantic information, such as lists of running
processes or active network connections, from low-level sources and detailed
knowledge of the OS \cite{rekall}.
The reader can refer to existing literature~\cite{jain2014sok} for
further information on this topic.

\subsection{Execution Manipulation}
Regardless of the level of introspection that every approach may offer, the
objective of DBI is not only to monitor the target under analysis, but
to possibly alter its behavior through the intentional manipulation of the
execution.

To this end, DBI frameworks may provide facilities through their APIs to add,
replace, or delete instructions by design. Moreover, other convenient facilities
include replacing specific function calls with user-defined functions, returning
prematurely, or modifying arguments and return values, for instance.
Nonetheless, program behavior and execution flow can also be altered through the
manipulation of the program state, which is usually accessible to the user.
Writing CPU registers or memory contents can lead the program to follow a
different path, by modifying the program counter or overwriting a branch
condition related value, for instance.

\section{Instrumentation Techniques}
\label{sec:tech}

The variety of events that a DBI system is able to instrument is subject to its
underlying techniques and implementation decisions. The most fundamental
observable events, called \textit{primitives}, include code execution and memory
accesses at different levels. Table~\ref{tb:dbi-events} summarizes the main
primitives in relation to the techniques that enable their instrumentation.

In contrast to previous work~\cite{d2019sok}, we do not include high-level
abstractions that are implementation dependent such as function calls,
since our goal is to express basic aspects of program execution.
Nevertheless, DBI systems frequently use symbol information or function
identification methods.

We also take a restrictive approach. We consider a technique to be capable of
instrumenting a primitive only if it is able by itself and if it is reasonably
applicable. For example, single-stepping (i.e., tracing every instruction
individually) cannot selectively instrument a specific instruction address, even
though it can single-step the entire program and examine every instruction
until that particular address. On the contrary, other techniques
can instrument specific instructions because the technique itself makes it
possible.

\begin{table*}[t]
    \small

    \caption{Instrumentation techniques and fundamental primitives. \cmark
    indicates full capability. \xmark means that the technique is not capable.
    \fullmoon indicates exceptional case or partial capability. Specific cases
    are detailed in their corresponding subsections.}\label{tb:dbi-events}

    \centering
    \setlength{\tabcolsep}{4pt}

    \begin{tabular}{l c c c c c c c}
        \toprule

        \textbf{Instrumentation Primitive} &
        \makecell{\textbf{JIT-DBT}} & 
        \makecell{\textbf{DPI}} & 
        \makecell{\textbf{SW} \\ \textbf{Breakpoints}} & 
        \makecell{\textbf{HW} \\ \textbf{Breakpoints}} & 
        \makecell{\textbf{Single-} \\ \textbf{Stepping}} & 
        \makecell{\textbf{Page} \\ \textbf{Faults}} & 
        \makecell{\textbf{CPU} \\ \textbf{Interp.}} \\

        \midrule

        \textit{Execution of a Single Instruction} & \cmark & \cmark & \cmark &
        \cmark & \xmark & \fullmoon & \cmark \\

        \textit{Execution of Instructions in Memory Range} & \cmark & \xmark &
        \xmark & \fullmoon & \xmark & \cmark & \cmark \\

        \textit{Execution of Every Instruction} & \cmark & \xmark & \xmark &
        \xmark & \cmark & \xmark & \cmark \\

        \textit{Execution of Specific Instruction Types (e.g., branches)} &
        \cmark & \xmark & \xmark & \xmark & \fullmoon & \xmark & \cmark \\

        \textit{Execution of Execution Block} & \cmark & \xmark & \xmark &
        \xmark & \xmark & \xmark & \xmark \\

        \textit{Block/Instruction Translation/Decoding} & \cmark & \xmark &
        \xmark & \xmark & \xmark & \xmark & \cmark \\

        \textit{Memory Address R/W} & \fullmoon & \xmark & \xmark & \cmark &
        \xmark & \fullmoon & \fullmoon \\

        \textit{Memory Range R/W} & \fullmoon & \xmark & \xmark & \fullmoon &
        \xmark & \cmark & \fullmoon \\

        \textit{Whole Memory R/W} & \fullmoon & \xmark & \xmark & \xmark &
        \xmark & \xmark & \fullmoon \\

        \bottomrule
    \end{tabular}
\end{table*}

\subsection{JIT-assisted Dynamic Binary Translation}
Dynamic Binary Translation (DBT) is a widespread technique extensively used in
technologies like emulation~\cite{bellard2005qemu} that also enables
instrumentation~\cite{henderson2014make,dolan2015panda,pyrebox}. It dynamically
translates code from one ISA to another. However, it is common that host and
guest architectures coincide~\cite{luk2005pin,bruening2004rio}. Moreover,
approaches may work on the native instruction set or use an intermediate
representation~\cite{nick2007valgrind}.

The process is assisted by a JIT compiler and a code cache. The original code is
not executed but usually compiled one block at a time, and stored in a
software-based code cache, which is the one that is actually executed. All the
instructions in this chain are executed one after the other unless some
interrupt or control flow instruction is reached.  This optimizes the
performance as it avoids switching between the emulator and the emulated code on
every single instruction.

These engines allow registering callbacks on the translation or
execution of code at different granularities: typically at execution block
level and instruction level, depending on the implementation. Further, some
systems may abstract memory accesses, exceptional control flows, or other CPU
and cache related events. These primitives can be used to emulate any kind of
hook by selectively inserting callbacks in JIT translated code (e.g., at a
certain memory address). Concrete memory accesses cannot be instrumented at
translation time, since in most cases the addresses are computed at
run-time. Instead, it has to examine executed instructions, or
directly monitor memory accesses in cases where the tool uses a
software Memory Management Unit (MMU). It is worth to note that JIT-DBT is the
only technique that naturally implements the execution block granularity level,
facilitating coarse-grained instrumentation and analysis.

\subsection{Dynamic Probe Injection}
The Dynamic Probe Injection (DPI) approach dynamically replaces instructions in the original
program with trampolines that jump into analysis code. To this end, instructions
are saved and replaced by an unconditional branch instruction. When this
trampoline is executed, the control flow changes to a pre-processing sequence
that pushes CPU registers into the stack to save the program state and
execute the user-defined code. When the callback returns, the program state is
restored, the instructions that were displaced are executed, and the control
flow returns to the original location~\cite{bernat2011anywhere,frida}.

This method is typically used for function-level hooking (i.e., intercept
calls to functions and execute code before or after the call), although it is
possible to target instructions at specific addresses to instrument. In
order to identify functions for instrumentation, DBI frameworks must rely on
symbol tables, debugging information, or other techniques such as 
static analysis to recover function boundaries.

\subsection{Traps and Interrupts}
Techniques commonly found in debuggers trigger traps or interrupts at specific
execution points, redirecting the execution to user code that will
typically have privileges over the target, as part of a debugger process
or a hypervisor. In this category we find software and hardware
breakpoints, watchpoints, single-stepping, and page-fault-based instrumentation.

\textbf{Software-based breakpoints} start by saving a specified instruction in
the program to replace it with an exception-triggering instruction. For
instance, in x86-64 platforms the exception-triggering instruction is usually a
special single-byte instruction (i.e., \texttt{INT 3}). After handling the
exception and executing callback code, the original instruction is written back
and executed~\cite{deng2013spider,lengyel2014drakvuf,rvmi}. Some approaches are
built around architecture-specific mechanisms such as the System Management Mode
(SMM). In that case, software breakpoints are implemented by replacing
instructions with a write to an Advanced Configuration and Power Interface
(ACPI) port (e.g., 0x2b on Intel) to trigger a System Management Interrupt (SMI)
and enter SMM~\cite{zhang2015malt}. On ARM, they are also implemented placing
Secure Monitor Call (SMC) instructions into the guest kernel to transfer control
to a hypervisor~\cite{proskurin2018hiding}. Similar to DPI, although this method
could theoretically instrument every instruction or a set of instructions in a
memory range, it would require setting and managing successive breakpoints on
each address, if possible.
  
\textbf{Hardware-based breakpoints} are implemented at CPU-level by using
dedicated debug registers (e.g., DR0-DR3 in the x86-64 architecture, or DBGBVR in
ARM). These registers are a limited resource and can store a breakpoint
address. When the program counter matches a value and a set of conditions are
met (i.e., determined by the DR7 register in x86-64, or DBGBCR in ARM), a debug
exception is triggered. Some architectures such as PowerPC may also include
hardware support for ranged breakpoints~\cite{gdb}.

\textbf{Watchpoints} are a special kind of breakpoint that breaks
when data is accessed rather than when some specific instruction is executed.
Hardware-assisted watchpoints are set leveraging
architecture-specific debug registers~\cite{proskurin2018hiding}, with support
for ranged watchpoints on architectures such as PowerPC~\cite{gdb}.  In
contrast, we do not contemplate purely software-based approaches since they
require single-stepping the program and monitoring every memory access to
compare the address against a list of watched addresses.  Nevertheless, several
optimizations have been proposed in that context~\cite{greathouse2012case}.

\textbf{Single-stepping} includes any technique that executes and handles one
step (i.e., an instruction) at a time, usually through mechanisms provided by
CPU architectures. In x86-64, the EFLAGS register contains the Trap Flag (TF),
that can be set to generate a debug exception on each executed
instruction~\cite{dinaburg2008ether,nguyen2009mavmm}. In addition, the Branch
Trap Flag (BTF) modifies the behavior of TF to only trigger exceptions on branch
instructions. Similarly, the Monitor Trap Flag (MTF) is a debugging feature
specific to Intel hardware virtualization extensions that causes VMExits on
instruction boundaries in VMX non-root operation (i.e., the guest
system)~\cite{deng2013spider,rvmi}. Further, when a performance counter
overflows an interrupt is triggered, enabling instruction-level stepping by
setting the value of a performance counter that measures all retired
instructions to the maximum value each time. The CPU architecture determines
stepping options by the kind of events that can be measured. For instance, it is
also possible to perform branch-level stepping in this way. The
triggered interrupts can be configured to transfer control to the hypervisor,
SMM, or the ARM secure domain~\cite{vogl2012using,zhang2015malt,ning2017ninja}.
On ARM, single-stepping can also be implemented placing successive SMC
instructions, taking advantage of the fixed-width
ISA~\cite{proskurin2018hiding}.

\textbf{Page Faults} operate at the memory page level and tamper with page
access bits to trigger page fault exceptions during execution, that
are handled by a previously set custom page fault handler. To this end, a page
that contains an instruction of interest can be set to
non-present~\cite{vasudevan2005stealth}. Further, when using Intel
virtualization extensions and hence Extended Page Tables (EPT), read, write, and
execute page access bits are independent of each other, providing complete
flexibility to trap the execution on different access types, triggering a
VMExit~\cite{yan2012v2e}. This method is not exclusive to code pages,
but is also applicable to data pages, allowing to implement a mechanism
similar to watchpoints. For instance, the privilege (i.e., supervisor bit) or
other access permission flags of the page can be manipulated to produce an
exception on every access~\cite{vasudevan2006cobra,deng2013spider}.

To monitor the execution of concrete instructions or specific memory
accesses, it is necessary to check on every page fault the responsible memory
address. Following the same principle, it is also possible to
monitor memory ranges not aligned to the page size for either read, write or
execution.

\subsection{CPU Interpretation}
The fetch-decode-execute cycle is the basic operational process the CPU
follows to execute instructions. This process is closely reproduced
in software by some emulators. In particular, interpretation-based emulators
work in a loop by fetching an instruction from memory, decoding
it to determine its semantics, fetching the particular operands, performing the
operations, and writing the result back. Instruction execution is usually
enclosed in an instruction-specific method that is called after its
decoding. This rigorous interpretation enables
instruction-level execution control and
instrumentation~\cite{lawton1996bochs,pyemu}.

Similar to JIT-DBT, this technique needs to examine each executed instruction
to check whether it is a memory operation to make memory accesses visible for
instrumentation. However, interpretation-based emulators usually emulate
virtual memory in software, making this task simpler.

\section{DBI Implementations}
\label{sec:map}

\begin{figure*}[ht]
\centering
\includegraphics[width=0.99\textwidth]{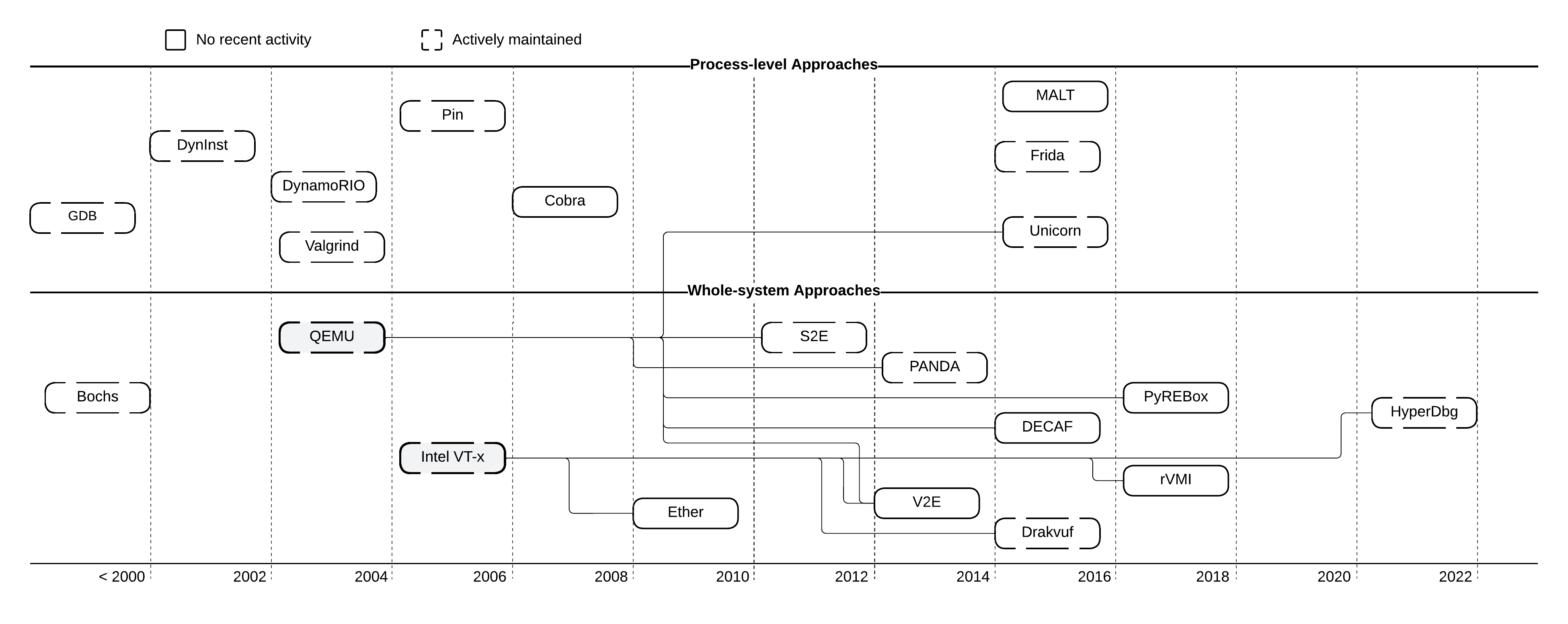}
\caption{Timeline of DBI tools.}
\label{fig:timeline}
\end{figure*}

In this section, we describe and relate existing DBI frameworks to their
implemented instrumentation techniques mapped to instrumentation primitives. To
compile the list of existing DBI frameworks, we have conducted a detailed review
of the literature and we have also included known tools in the system security
community, of which most are widely used both in industry and academia. Please
note that we have excluded tools that mix or are built upon other DBI systems
(e.g., PEMU~\cite{zeng2015pemu}), and tools whose details are not publicly
available.  Similarly, we do not include tools with a more specific purpose such
as TinyInst~\cite{tinyinst} (whose goal is to instrument specific modules in a
process).

To depict the temporal evolution of proposed DBI frameworks,
Figure~\ref{fig:timeline} presents a timeline of DBI solutions. We lay out the
tools according to their first published work or first public release. Please
note that, even though some tools came out early on, they have been further
developed with new features in posterior versions, have published additional
works, and continue to be widely used and actively maintained. We also include
QEMU and hardware-assisted virtualization technologies (i.e., Intel VT-x)
because they have been a fundamental component in DBI, and several approaches
depend on them.

\subsection{Process-Level Approaches}
\label{ss:singleprocess}

\begin{table*}[ht!]
    \footnotesize

    \caption{Process-level DBI tools mapped to instrumentation techniques and
    primitives. J stands for JIT-DBT, D for dynamic probe injection, SB for
    software breakpoints, HB for hardware breakpoints, SS for single-stepping,
    PF for page faults, and M for software emulated memory. \xmark denotes not
    capable.}\label{tb:sp-tools}

    \centering

    \begin{tabular}{l c c c c c c c c c c}
        \toprule

        {} & \multicolumn{6}{c}{\textbf{JIT-DBT}} &
        \multicolumn{2}{c}{\textbf{DPI}} &
        \multicolumn{2}{c}{\textbf{Others}} \\

        \cmidrule(lr{4pt}){2-7}
        \cmidrule(lr{4pt}){8-9}
        \cmidrule(lr{4pt}){10-11}

        \textbf{Instrumentation Primitive} &
        \rotatebox{80}{\textit{Pin}~\cite{luk2005pin}} &
        \rotatebox{80}{\textit{DynamoRIO}~\cite{bruening2004rio}} &
        \rotatebox{80}{\textit{Valgrind}~\cite{nick2007valgrind}} &
        \rotatebox{80}{\textit{Frida Stalker}~\cite{frida}} &
        \rotatebox{80}{\textit{Unicorn}~\cite{unicorn}} &
        \rotatebox{80}{\textit{Cobra}~\cite{vasudevan2006cobra}} &
        \rotatebox{80}{\textit{DynInst}~\cite{bernat2011anywhere}} &
        \rotatebox{80}{\textit{Frida Interceptor}~\cite{frida}} &
        \rotatebox{80}{\textit{MALT}~\cite{zhang2015malt}} &
        \rotatebox{80}{\textit{GDB}~\cite{gdb}} \\

        \midrule

        \textit{Execution of Single Instruction} & \textbf{J} & \textbf{J} &
        \textbf{J} & \textbf{J} & \textbf{J} & \textbf{J} & \textbf{D} &
        \textbf{D} & \textbf{SB, HB} & \textbf{SB, HB, PF} \\

        \textit{Execution of Instructions in Memory Range} & \textbf{J} &
        \textbf{J} & \textbf{J} & \textbf{J} & \textbf{J} & \textbf{J} & \xmark
        & \xmark & \xmark & \textbf{HB*, PF} \\

        \textit{Execution of Every Instruction} & \textbf{J} & \textbf{J} &
        \textbf{J} & \textbf{J} & \textbf{J} & \textbf{J} & \xmark & \xmark &
        \textbf{SS} & \textbf{SS} \\

        \textit{Execution of Specific Instruction Types (e.g., branches)} &
        \textbf{J} & \textbf{J} & \textbf{J} & \textbf{J} & \textbf{J} &
        \textbf{J} & \textbf{D*} & \xmark & \textbf{SS} & \xmark \\

        \textit{Execution of Execution Block} & \textbf{J} & \textbf{J} &
        \textbf{J} & \textbf{J} & \textbf{J} & \textbf{J} & \textbf{D*} & \xmark
        & \xmark & \xmark \\

        \textit{Block/Instruction Translation/Decoding} & \textbf{J} &
        \textbf{J} & \textbf{J} & \textbf{J} & \xmark & \textbf{J} & \xmark &
        \xmark & \xmark & \xmark \\

        \textit{Memory Address R/W} & \textbf{J} & \textbf{J} & \textbf{J} &
        \textbf{J, PF} & \textbf{M} & \textbf{PF} & \textbf{D*} & \textbf{PF} &
        \xmark & \textbf{HB, PF} \\

        \textit{Memory Range R/W} & \textbf{J} & \textbf{J} & \textbf{J} &
        \textbf{J, PF} & \textbf{M} & \textbf{PF} & \textbf{D*} & \textbf{PF} &
        \xmark & \textbf{HB*, PF} \\

        \textit{Whole Memory R/W} & \textbf{J} & \textbf{J} & \textbf{J} &
        \textbf{J} & \textbf{M} & \xmark & \xmark & \xmark & \xmark & \xmark \\

        \bottomrule
        \addlinespace[0.1cm]
        \multicolumn{11}{l}{\footnotesize * Partial capability, specific to a
        CPU architecture, or tool-specific implementation.} \\
    \end{tabular}
\end{table*}

Table~\ref{tb:sp-tools} shows the implemented instrumentation techniques within
process-level tools that enable the instrumentation of different primitives.

Early implementations of JIT-DBT (i.e., DynamoRIO, Valgrind, Pin) made
fine-grained instrumentation practical. Tools that implement JIT-DBT share
address space with the target program and provide efficient runtime
instrumentation at a wide range of granularities. Each tool implements different
designs, translation strategies and optimizations, and operate on native code.
The exception is Valgrind, which translates through the VEX intermediate
representation and enables heavyweight analyses such as in-depth memory analysis
and taint tracking.

Posterior implementations focused on additional aspects of DBI. For instance,
Cobra was developed as a kernel module to enable the analysis of localized user
and kernel code, also improving stealth. Frida is oriented to portable,
user-friendly, cross-platform runtime instrumentation, with a strong focus on
mobile. It injects a JavaScript engine into the target process to enable DBI
through JavaScript code. Unicorn is based on the CPU emulation component of
QEMU, which enables cross-platform, lightweight emulation and instrumentation of
arbitrary binary code such as firmware blobs. Since QEMU emulates virtual
memory, it makes it possible to intercept memory accesses at TLB-level.

\begin{table*}[htp]
    \footnotesize

    \caption{Whole-system DBI systems mapped to instrumentation techniques and
    primitives. J stands for JIT-DBT, D for dynamic probe injection, SB for
    software breakpoints, SS for single-stepping, PF for page faults, CI for CPU
    interpretation, and M for software emulated memory. \xmark denotes not
    capable.}\label{tb:ws-tools}

    \centering

    \begin{tabular}{l c c c c c c c c c c}
        \toprule

        {} & \multicolumn{6}{c}{\textbf{Emulation}} &
        \multicolumn{4}{c}{\textbf{Hypervisor}} \\

        \cmidrule(lr{4pt}){2-7}
        \cmidrule(lr{4pt}){8-11}

        \textbf{Observable Event} &
        \rotatebox{80}{\textit{DECAF}~\cite{henderson2014make}} &
        \rotatebox{80}{\textit{PANDA}~\cite{dolan2015panda}} &
        \rotatebox{80}{\textit{V2E}~\cite{yan2012v2e}} &
        \rotatebox{80}{\textit{S2E}~\cite{chipounov2011s2e}} &
        \rotatebox{80}{\textit{PyREBox}~\cite{pyrebox}} &
        \rotatebox{80}{\textit{Bochs}~\cite{lawton1996bochs}} &
        \rotatebox{80}{\textit{Ether}~\cite{dinaburg2008ether}} &
        \rotatebox{80}{\textit{Drakvuf}~\cite{lengyel2014drakvuf}} &
        \rotatebox{80}{\textit{rVMI}~\cite{rvmi}} &
        \rotatebox{80}{\textit{HyperDbg}~\cite{karvandi2022hyperdbg}} \\

        \midrule

        \textit{Execution of Single Instruction} & \textbf{J} & \textbf{J} &
        \textbf{J} & \textbf{J} & \textbf{J} & \textbf{CI} & \xmark &
        \textbf{SB, PF} & \textbf{SB} & \textbf{SB, PF, D*} \\

        \textit{Execution of Instructions in Memory Range} & \textbf{J} &
        \textbf{J} & \textbf{J} & \textbf{J} & \textbf{J} & \textbf{CI} & \xmark
        & \textbf{PF} & \xmark & \xmark \\

        \textit{Execution of Every Instruction} & \textbf{J} & \textbf{J} &
        \textbf{J} & \textbf{J} & \textbf{J} & \textbf{CI} & \textbf{SS} &
        \xmark & \textbf{SS} & \textbf{SS} \\

        \textit{Execution of Specific Instruction Types (e.g., branches)} &
        \textbf{J} & \textbf{J} & \textbf{J} & \textbf{J} & \textbf{J} &
        \textbf{CI} & \xmark & \xmark & \xmark & \textbf{PF*} \\

        \textit{Execution of Execution Block} & \textbf{J} & \textbf{J} &
        \textbf{J} & \textbf{J} & \textbf{J} & \xmark & \xmark & \xmark & \xmark
        & \xmark \\

        \textit{Block/Instruction Translation/Decoding} & \textbf{J} &
        \textbf{J} & \xmark & \textbf{J} & \xmark & \textbf{CI} & \xmark &
        \xmark & \xmark & \xmark \\

        \textit{Memory Address R/W} & \textbf{M} & \textbf{M} & \textbf{M} &
        \textbf{M} & \textbf{M} & \textbf{M} & \xmark & \textbf{PF} &
        \textbf{HB} & \textbf{PF}\\

        \textit{Memory Range R/W} & \textbf{M} & \textbf{M} & \textbf{M} &
        \textbf{M} & \textbf{M} & \textbf{M} & \xmark & \textbf{PF} & \xmark &
        \textbf{PF}\\

        \textit{Whole Memory R/W} & \textbf{M} & \textbf{M} & \textbf{M} &
        \textbf{M} & \textbf{M} & \textbf{M} & \textbf{PF*} & \xmark & \xmark &
        \xmark \\

        \bottomrule
        \addlinespace[0.1cm]
        \multicolumn{10}{l}{\footnotesize * Partial capability or tool-specific
        implementation.} \\
    \end{tabular}
\end{table*}

In contrast to JIT-DBT, certain approaches do not share the same address space
and operate from an external point, i.e., from a different process or a
different physical machine (i.e., MALT). GDB implements standard debugger
features like breakpoints and single-stepping. In GDB, page-permission-based
traps are not implemented by default. However, the user can call arbitrary
functions in the debugged process space. This enables their implementation by
calling \texttt{mprotect} to modify page permissions and trigger page fault
exceptions on access. Resulting page faults are handled by the debugger process,
permissions are restored back in the faulting page, the instruction is executed
in single-step, and page permissions are adjusted again before the execution
continues. GDB also implements ranged hardware breakpoints and watchpoints for
PowerPC embedded processors.

DynInst introduced dynamic probe injection in DBI. It implements several static
analyses to identify and build abstractions that enable the instrumentation of
more complex primitives using DPI. To illustrate, analysis tools can build the
control-flow graph of specific functions and instrument basic block boundaries
(different from execution blocks in JIT-DBT). Similarly, it also enables the
instrumentation of load and store instructions in particular functions.

MALT emphasized the role of hardware features in improving DBI transparency by
combining the SMM and performance counters. It introduced a new method to
implement debugger-related instrumentation techniques by leveraging performance
counter events to trigger an SMI.

\subsection{Whole-System Approaches}
\label{ss:whole}

Table~\ref{tb:ws-tools} shows the implemented instrumentation techniques within
whole-system tools that enable the instrumentation of primitives. These tools
are built upon hardware virtualization, emulation, or a combination of both
technologies.

Bochs introduced full-system emulation for x86 CPUs and whole-system
instrumentation. It is an interpretation-based emulator, i.e., emulation logic
is implemented in one big decode loop, which closely models the
fetch-decode-execute actions of the CPU, enabling fine-grained instrumentation.

At present, JIT-DBT is the most common approach in emulation-based tools, most
of which are based on QEMU's whole-system emulator. However, significant
differences exist between tools and their objectives. S2E, PANDA and DECAF
enable fine-grained instrumentation in addition to different complex analyses
such as symbolic execution in S2E, deterministic execution record and replay in
PANDA, and taint analysis in DECAF, among others. Exceptionally, V2E is a hybrid
approach that records the execution at page-level in KVM, and replays it in a
modified version of TEMU~\cite{yin2010temu}, taking advantage of JIT-DBT to
perform fine-grained instrumentation. Further on, PyREBox combined QEMU-based
whole-system instrumentation with Python scripting, improving ease-of-use for
system-wide reverse engineering and analysis.

Other whole-system tools are built upon an hypervisor component such as Xen or
KVM \cite{lengyel2014drakvuf,rvmi,dinaburg2008ether}. Hardware virtualization
extensions allowed tools to access memory and CPU state faster and
transparently. Ether spearheaded the transition from intrusive tools to stealthy
hypervisor-based DBI by leveraging hardware-assisted virtualization. It uses the
Trap Flag and relies on shadow page tables, which makes it possible to also
intercept memory writes by removing write permissions from all pages. However,
it has limited performance and capabilities since shadow page tables were less
efficient than the later released EPT.

Drakvuf, rVMI and HyperDbg leverage EPT to implement instrumentation that
transfers control to the hypervisor, achieving greater stealth and performance.
Drakvuf also implements memory access traps by removing the read/write or
execution permissions from target pages at EPT level, so the page fault is not
triggered in the guest. Moreover, HyperDbg has a strong focus on stealth,
offering several distinctive features. It combines multiple instrumentation
techniques making maximal use of EPT, including breakpoints, single-stepping via
Monitor Trap Flag, page faults, and dynamic probes for greater flexibility and
transparent analysis.

\subsection{Other Execution Events}
\label{ss:further}

\begin{table*}[htp]
    \footnotesize

    \caption{Other visible events derived from execution, 
             OS, or hardware operation.}\label{tb:execevents}

    \centering

    \setlength{\tabcolsep}{3pt}

    \begin{tabular}{l c c c c c c c c c c c c c c c c c c}
        \toprule

        {} & \multicolumn{8}{c}{\textbf{Process-Level}} &
        \multicolumn{10}{c}{\textbf{Whole-System}} \\

        \cmidrule(lr{4pt}){2-9}
        \cmidrule(lr{4pt}){10-19}

        \textbf{Observable Event} &

        \rotatebox{80}{\textit{Pin}~\cite{luk2005pin}} &
        \rotatebox{80}{\textit{DynamoRIO}~\cite{bruening2004rio}} &
        \rotatebox{80}{\textit{Valgrind}~\cite{nick2007valgrind}} &
        \rotatebox{80}{\textit{Frida}~\cite{frida}} &
        \rotatebox{80}{\textit{DynInst}~\cite{bernat2011anywhere}} &
        \rotatebox{80}{\textit{Cobra}~\cite{vasudevan2006cobra}} &
        \rotatebox{80}{\textit{Unicorn}~\cite{unicorn}} &
        \rotatebox{80}{\textit{GDB}~\cite{gdb}} &

        \rotatebox{80}{\textit{DECAF}~\cite{henderson2014make} } &
        \rotatebox{80}{\textit{PANDA}~\cite{dolan2015panda}} &
        \rotatebox{80}{\textit{S2E}~\cite{chipounov2011s2e}} &
        \rotatebox{80}{\textit{PyREBox}~\cite{pyrebox}} &
        \rotatebox{80}{\textit{Ether}~\cite{dinaburg2008ether}} &
        \rotatebox{80}{\textit{Drakvuf}~\cite{lengyel2014drakvuf}} &
        \rotatebox{80}{\textit{rVMI}~\cite{rvmi}} &
        \rotatebox{80}{\textit{V2E}~\cite{yan2012v2e}} &
        \rotatebox{80}{\textit{Bochs}~\cite{lawton1996bochs}} &
        \rotatebox{80}{\textit{HyperDbg}~\cite{karvandi2022hyperdbg}} \\

        \midrule

        \textbf{Interrupts \& Signals} & \cmark & \cmark & \cmark &
        \cmark & \cmark & \xmark & \cmark & \cmark & \xmark & \cmark & \cmark &
        \xmark & \xmark & \cmark & \xmark & \cmark & \cmark & \cmark \\

        \textbf{Process Creation \& Exit} & \cmark & \cmark & \xmark & \xmark &
        \cmark & \xmark & \xmark & \cmark & \cmark & \cmark & \cmark & \cmark &
        \xmark & \cmark & \xmark & \cmark & \xmark & \cmark \\

        \textbf{Thread Start \& Exit} & \cmark & \cmark & \cmark & \xmark &
        \cmark & \xmark & \xmark & \cmark & \xmark & \cmark & \xmark & \xmark &
        \xmark & \xmark & \xmark & \xmark & \xmark & \xmark\\
        
        \textbf{System Call} & \cmark & \cmark & \cmark & \xmark & \xmark &
        \cmark & \cmark & \cmark & \xmark & \cmark & \cmark & \xmark & \cmark &
        \cmark & \xmark & \xmark & \xmark & \cmark \\

        \textbf{Module Load \& Unload} & \cmark & \cmark & \xmark & \xmark &
        \cmark & \xmark & \xmark & \cmark & \cmark & \xmark & \cmark & \cmark &
        \xmark & \cmark & \xmark & \cmark & \xmark & \xmark\\

        \textbf{Task Switch} & \xmark & \xmark & \xmark & \xmark & \xmark &
        \xmark & \xmark & \xmark & \xmark & \cmark & \cmark & \cmark & \cmark &
        \cmark & \cmark & \xmark & \cmark & \xmark \\

        \textbf{Network I/O} & \xmark & \xmark & \xmark & \xmark & \xmark &
        \xmark & \xmark & \xmark & \cmark & \cmark & \xmark & \cmark & \xmark &
        \cmark & \xmark & \cmark & \xmark & \cmark \\

        \textbf{HDD I/O} & \xmark & \xmark & \xmark & \xmark & \xmark & \xmark &
        \xmark & \xmark & \xmark & \cmark & \xmark & \xmark & \xmark & \cmark &
        \xmark & \cmark & \xmark & \cmark \\
        
        \textbf{TLB Operation} & \xmark & \xmark & \xmark & \xmark & \xmark &
        \xmark & \xmark & \xmark & \cmark & \xmark & \cmark & \cmark & \xmark &
        \xmark & \xmark & \xmark & \cmark & \xmark \\

        \bottomrule
    \end{tabular}
\end{table*}

During execution, the execution flow can be altered by exceptions and
interrupts. Some of these come from the interaction with external devices such
as the hard-disk drive, the keyboard or the network card. The OS is usually
responsible for their management. Some DBI tools have visibility on these
events, subject to implementation decisions and symbol availability.
Table~\ref{tb:execevents} shows the relation between DBI tools and the most
common event types within this category. We conservatively take into
consideration current implementations and default plugins, which usually expose
these events through their APIs.

Process-level and whole-system tools are implemented with different goals in
mind. Therefore, they observe events from different perspectives and some events
may not be captured by a particular category. Specifically, as process-level
approaches focus on a single program, they track events that take place
in the content of the program under analysis, while whole-system DBI observes
events in a system-wide fashion. As an example, process-level tools can
intercept the creation of child processes originating from the target.
Whole-system approaches can instead monitor the creation of every process.
Likewise, process-level approaches capture module loads and unloads in the
process (i.e., shared libraries), whereas whole-system solutions also cover
kernel modules or drivers.

Existing tools adopt different strategies to handle interrupts. Process-level
tools let the user subscribe to exception events (e.g., division by zero) or
Linux-specific signal events, depending on the implementation. Valgrind and
DynInst only support signals while Frida supports exceptions.
The Unicorn Engine is rather exceptional since it is a CPU emulator and thus,
although it provides callback functions for exceptions and \texttt{INT}
instructions, it is the user's job to emulate the entire exception handling
process. Whole-system approaches do not usually provide primitives to monitor
signals or exceptions, but they have full visibility on the system, making it
possible to create these abstractions by using VMI techniques.

\section{Evaluation}
\label{sec:eval}

In this section, we detail the conducted experiments to measure and
evaluate the performance of the different instrumentation techniques.
We seek to understand the behavior of instrumentation techniques in different
scenarios, such as targeting different primitives, or different
event frequency.

\subsection{Instrumentation Techniques in Practice}
\label{ss:techniques}
Certain techniques such as JIT-DBT are complex and, as a consequence, tools
implementing the same technique may show performance differences due to their
particular implementation or design choices, among other factors. Therefore,
our main goal is to evaluate instrumentation techniques and not their particular
implementations or available tools.

However, in order to conduct these experiments we had to select a set of tools
as reference implementations of the techniques. We prioritized publicly
available, actively maintained projects written in C/C++ with a large community.
Additionally, we give preference to tools that do not introduce excessive
unrelated overhead (i.e., to enable complex analyses), so that they do not
distort the real cost of instrumentation.

The implementation language of both the tool and the instrumentation code have a
significant performance impact. For instance, a tool written in C/C++ that calls
a Python snippet needs to enter the Python run-time on every event, imposing a
high overhead. This overhead is derived from a design choice and not by the
instrumentation technique itself. All the tools we selected are fully
implemented in C/C++. We conducted our experiments on the Linux platform for the
x86-64 architecture, which should be supported by the reference tools. While
there are DBI tools that support other architectures, they follow similar
techniques and principles.  Also, it would not be appropriate to compare x86-64
implementations against other architectures.

In this way, we selected tools to cover every possible instrumentation technique
for process-level and whole-system. Each of these tools
implement one or more of the instrumentation techniques described in Section
\textsection\ref{sec:tech}. 

Specifically, we used the following process-level technique
implementations: the JIT-DBT engine Pin (3.17), the DPI tool Dyninst (10.2.1),
and the GDB (9.2) debugger for software breakpoints, hardware
breakpoints, and page-fault-based instrumentation. While GDB is a debugger and
not a DBI tool on its own, it implements several instrumentation
techniques and satisfies the definition of DBI in Section
\textsection\ref{sec:bblocks}. Also note that we configured GDB to prevent all
output to avoid introducing unnecessary overhead.

As whole-system implementations, we selected: the
interpretation emulator Bochs (2.6.11), and the latest available versions of
the JIT-DBT emulator PANDA and the hypervisor-based tool Drakvuf. These
tools cover a wide range of techniques, described in Section \textsection\ref{sec:tech}.
In cases where callbacks can be inserted in different programming languages
(e.g., PANDA supports plugins in Python), we used the C/C++ interface.

Note that in the rest of this section, in order to make a distinction between
the different techniques implemented by the same tool, we append the acronym of
the technique at the end of the tool name (i.e., \texttt{gdb\_sb},
\texttt{gdb\_hb}, \texttt{gdb\_ss}, \texttt{gdb\_pf}, \texttt{drakvuf\_sb}, and
\texttt{drakvuf\_pf}).

\subsection{Technique Evaluation}

Our goal is to analyze current instrumentation
techniques, evaluating their performance when implementing different
instrumentation primitives. It is essential to minimize the impact of specific
implementations. These rules out complex test cases that use multiple
instrumentation options, or high demanding tasks such as memory
profiling or information flow tracking, because tool-specific optimizations
would come into play. Under that premise, our experiments are guided by and
reflect basic instrumentation capabilities. 

We did not cover the execution of specific instruction types, execution of
execution blocks, or code translation primitives. While these are useful
primitives in many contexts and applications, reporting their performance
statistics would not add any additional value since they are either based on the
same principles as other primitives in the techniques, or not applicable to more
than one technique. For instance, execution block primitives are exclusive to
JIT-DBT by design.

It is worth noting that some instrumentation techniques were really demanding,
leading to overly long execution times and making the measurement impractical.
To overcome this difficulty, we empirically set through experimentation an event
count threshold for these cases, thus keeping the execution feasible: 2 million
instructions when single-stepping, and 200,000 page faults when using a
page-fault-based approach.

\noindent \textbf{No instrumentation (\textit{none}).}
For every tool, we run the target under the DBI framework with no
instrumentation besides the necessary to measure performance. This 
serves as a baseline to understand the base overhead introduced by every tool.

In Bochs, identifying the process start and end requires using the
instruction-level callback, which introduces excessive overhead for this
experiment. To overcome this limitation, we compiled Bochs without any
instrumentation enabled, and connected an agent in the guest system with the
host through a network socket in order to properly detect the beginning and end
of the program without adding any expensive overhead.

\noindent \textbf{Instrument execution of a single instruction
(\textit{exec\_single}).}
This primitive can be implemented in different ways. JIT-DBT engines can decide
whether to instrument an instruction at translation time based on the address
being translated. This is the case of Pin and PANDA. DPI techniques can insert
trampolines at specific addresses -- we implemented this approach in Dyninst.
Software and hardware breakpoints trigger an exception at a specific
code address. A similar behavior is possible by removing
the execution permission of the whole memory page containing the address. We
implemented these three approaches in GDB and Drakvuf (except hardware
breakpoints for the latter). Finally, when an interpreter runs every instruction
one by one, it can call a callback before instruction
execution. We tested this approach with Bochs.

\noindent \textbf{Instrument execution of instructions in a memory range
(\textit{exec\_range}).}
Similar to previous cases, JIT-DBT engines can check at instruction
translation whether the current program counter value is within the range of a
specified memory range. We tested this with Pin and PANDA. This technique is
also possible through page faults. When the memory range is not aligned to
page boundaries, it is necessary to check (as part of the instrumentation code)
whether the faulting address is within the specified range. We implemented this
technique in GDB, and Drakvuf. As in the previous case, the Bochs
interpreter can check the address instruction by instruction.

\noindent \textbf{Instrument every executed instruction (\textit{exec\_all}).}
Not every instrumentation technique has the capacity to instrument every
possible instruction in a straightforward manner. This primitive can be
implemented either by using single-stepping (that will force an exception on
every instruction) or by monitoring every instruction execution under a JIT-DBT
engine or an interpreter. We tested the first technique with GDB, and the second
one with Pin for process-level, and PANDA and Bochs for whole-system
instrumentation.

\noindent \textbf{Instrument memory read/writes to a memory address
(\textit{rw\_single}).}
We implemented this primitive using different techniques. Hardware breakpoints
trap the execution on memory accesses. It is also possible to mimic
this behavior at page granularity by removing read and write permissions of 
memory pages. We implemented both techniques in GDB. For JIT-DBT
and interpretation, the only way is to monitor every memory
access to check whether the accessed address matches. Due to indirect memory
addressing, many addresses are computed at run-time, and it is not
possible to instrument this primitive at translation time. We implemented this
approach in Pin, PANDA, and Bochs.

\noindent \textbf{Instrument memory read/writes to a memory range
(\textit{rw\_range}).}
We implemented this instrumentation primitive by using the same techniques
described previously. We implemented them in Pin, GDB (only using page faults),
PANDA, and Bochs. In a similar vein, we included Drakvuf 
by setting a memory access trap for read and write accesses at
the specified memory pages at the EPT level.

\noindent \textbf{Instrument every memory read/write made by the program
(\textit{rw\_all}).}
This final primitive was only implemented by using JIT-DBT and interpretation,
monitoring every single memory access at run-time. We implemented this test in
Pin, PANDA, and Bochs.

\subsection{Testing Benchmarks}
For our experiments, we used the SPEC CPU2006 suite, which contains
diverse programs written in C/C++, integer or floating-point intensive, and
with varying execution times, instruction footprints and cache sensitivity.
We discarded those that did not compile or execute correctly in our system, or
those with excessively long execution times. We also used the
intermediate size workload.

The SPEC suite defines multiple executions in a single benchmark for some
programs, i.e., using different command-line options. We present these cases
individually. Hence, we finally used a total of 16 benchmarks
from 10 unique binaries.

For the experiments that require specific memory addresses to instrument, we
manually searched for instructions and memory addresses that were respectively
executed or accessed a significant number of times in each binary. To do so, we
iteratively debugged each program monitoring candidate addresses each round.

\subsection{Evaluation Methodology and Metrics}

\begin{figure}[t]
\centering
\includegraphics[width=0.48\textwidth]{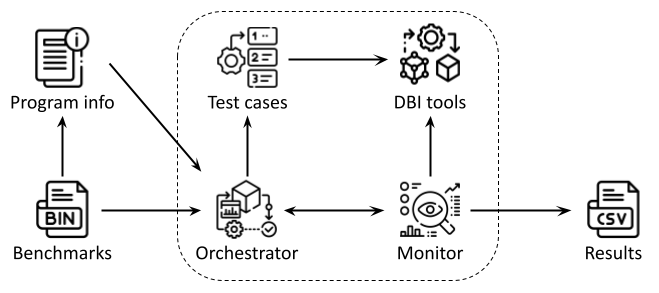}
\caption{Overview of the DBI evaluation framework.}
\label{fig:eval_overview}
\end{figure}

To manage and automate experiment execution, and enable further analyses, we
developed an evaluation tool that is depicted in Figure~\ref{fig:eval_overview}.
The proposed evaluation tool takes the selected benchmark binaries as input,
along with the necessary program-specific information for each benchmark to run
the experiments. This information includes command-line arguments,
instrumentation points, and other relevant memory addresses. Then, it executes
every test in every tool and every SPEC program in the evaluation set, recording
execution information. We executed each benchmark 10 times to minimize the
effect of system irregularities on the results.

To keep consistency, we use the execution of \texttt{main} as the
starting point and the execution of \texttt{exit} as the ending point. When the
execution reaches these, the tool notifies the framework and retrieves time
and status information. Since each tool provides different methods to identify
the start and end points, we selected the least intrusive method to avoid adding
overhead. Note that since we measure from \texttt{main} to \texttt{exit},
we avoid measuring time that tools would need to repeat their initialization
(e.g., creating a code cache) on every test case, minimizing the impact of the
tools on the measured executions.

Regarding Pin, Dyninst, and GDB; we use standard tool features such as hooks
and breakpoints to catch the execution of \texttt{main} and
\texttt{exit}. In Drakvuf, we place a breakpoint in the
\texttt{do\_syscall} kernel function to trap system calls. This allows us to
watch the start and end of the specific processes. To this end, we only need
to place a breakpoint in \texttt{main} when it started. To identify the starting
point in PANDA, we use the instruction translation callback to only instrument
instructions whose program counter matches the specific \texttt{main} address.
Whenever it is executed, we compare the next 20 bytes from \texttt{main} to
prevent misidentifying the start. To identify the ending point, we look into TCG
blocks before host code generation to find syscalls. In the case that a syscall
is found (i.e., opcode 0F 05), we insert a function call to verify whether it is
an actual exit called by our process (i.e., 0x3c or 0xe7 syscall number stored
in RAX register). Identification in Bochs is conceptually similar to PANDA.
Before instruction execution, we verify both if the program counter and the
following 20 bytes match for the starting point, and whether the current
instruction is an exit syscall for the ending point. Unlike PANDA, callbacks
cannot be disabled and there is no less intrusive method.

We obtain diverse information about the process from each execution. Every time
the orchestrator is notified, it retrieves wall-clock time from a system-wide
real-time clock and cpu-time from a high-resolution per-process timer from the
CPU. In addition, it creates a copy of \texttt{/proc/pid/stat} and
\texttt{/proc/pid/status}. When the process ends, it computes the difference
between the starting and ending points. 

For process-level, we use cpu-time as a reference time for our stats, except for
DynInst and GDB, which are executed as a separate process (i.e, a tracer process
that controls the execution of a child process). In these two cases we consider
a combination of the time spent by the child process (i.e., the monitored
process) and the cpu-time of the process corresponding to the instrumentation
tool.
Additionally, we obtained the event count for the instrumented primitive. The
event is defined by the executed experiment, e.g., for \textit{exec\_all}: the
total number of executed instructions, for \textit{rw\_single}: the access count
to the specified memory address, and so on.

For whole-system, we recorded wall-clock time from process start until
the end. We found that this is the most consistent way to evaluate the
approaches under the same conditions, considering that some of them run emulated
clocks inside the guest, while others may use the host's clock. Some run
as user-mode (e.g., emulators), while others run partially inside a VM,
partially in the hypervisor and partially in the host.

We conducted all the experiments on an Intel Core i7-8700 machine with 16GB of
RAM, running Ubuntu 20.04 with kernel version 5.8.0-63. For the guest machine,
we used an Ubuntu 20.04 cloud image with kernel version 5.4.0-54 and 4GB of RAM.
We also disabled ASLR, kASLR and cpu mitigations in both host and guest
machines.

\subsection{Process-level Techniques Results}
\label{ss:results_process_level}
Table~\ref{tbl:single_process} in Appendix~\ref{appendix:res} shows the results
of our process-level experiments for Pin, GDB software breakpoints
(\textit{gdb\_sb}), GDB hardware breakpoints (\textit{gdb\_hb}), GDB
single-stepping (\textit{gdb\_ss}), GDB page faults (\textit{gdb\_pf}), and
Dyninst. For every combination of experiment, tool, benchmark, and type of
instrumentation, we show the average and standard deviation.

\begin{figure}[h]
\centering
\includegraphics[width=0.48\textwidth]{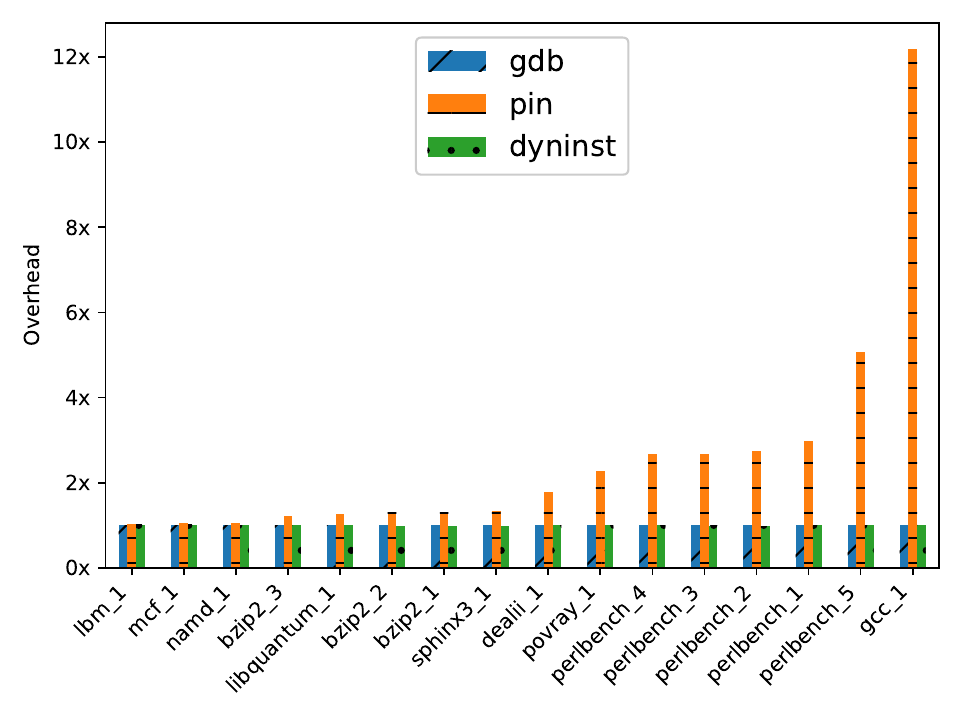}
\caption{Base overhead of process-level DBI systems.}
\label{fig:single_overhead}
\end{figure}

Figure~\ref{fig:single_overhead} shows the base performance overhead of each
tool. Debugging-based tools use operating-system
provided mechanisms to attach to the debugged process. When no instrumentation
is done, the overhead of this kind of tools is minimal or nonexistent.
Trampoline based tools install trampolines --- or code detours --- in a given
address, generating very little overhead as well. In contrast, dynamic
compilation tools such as Pin, need to read and analyze the original code to give
the opportunity to insert instrumentation routines before compilation.  This
process implies an overhead of up to 12 times the reference execution time on
the native system even when no instrumentation is performed.
Figure~\ref{fig:single_overhead} also shows how the overhead remains constant
for GDB and Dyninst, which execute the code natively, while some programs have
different overheads for Pin. The reason is that the nature of the program and
its execution also affect JIT-DBT, which may increase the time spent in branch
resolution or register allocation~\cite{uh2006analyzing}. In this case,
instruction footprint and execution time are responsible for the increased
overhead~\cite{luk2005pin}, especially in \texttt{gcc} benchmark. Having a
large instruction footprint and a short execution time results in insufficient
code reuse to compensate the compilation cost.

\begin{figure}[h]
\centering
\includegraphics[width=0.48\textwidth]{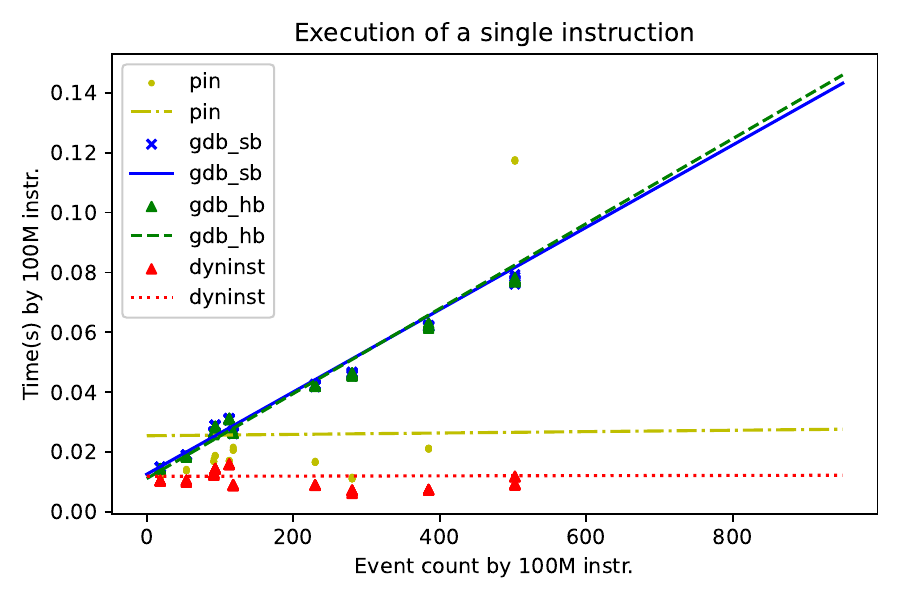}
\caption{Time required to execute 100 million instructions by number of single
instruction execution events. Trend lines represent the linear regression of
the plotted points.}
\label{fig:one_instruction_single}
\end{figure}

Even though binary rewriting inflicts a significant overhead with respect to
native execution, this might be tolerable under certain circumstances. For
instance, debuggers use software and hardware breakpoints or page faults to
trigger interrupts. Every method that triggers an interrupt
is followed by a context switch. The OS will analyze that
interrupt and redirect the execution to the corresponding handler. This process
implies the execution of many operations (e.g., saving the interrupted process's
context, context switching, transitioning from Ring 3 to Ring 0, and back to
Ring 3). Hence, while the number of instrumented events increases, the overhead
of JIT-DBT techniques is compensated by how efficiently those approaches can
instrument the code (e.g., no context-switch is needed).

Figure~\ref{fig:one_instruction_single} shows the normalized time required to
execute a single instrumented instruction for several programs with a number of
instrumentation events (i.e., the selected instruction is executed a different
number of times in each benchmark program). The X axis shows the number of
instrumented events every 100M instructions, whereas the Y axis shows the time
to execute those 100M instructions. Despite the high variability in
the case of Pin (discussed previously), all the cases show a linear overhead growth as the number of
instrumented events per 100M instructions increases. However, the growth rate of
GDB is much higher and exceeds the overhead of Pin around 100 events per 100M
instructions. Despite this number is indicative and depends on many factors, Pin
clearly outperforms GDB as soon as it needs to instrument a few events. Dyninst
performs even better than Pin as it benefits from both not needing to switch
context and not needing to implement binary rewriting.

\begin{figure}[h]
\centering
\includegraphics[width=0.48\textwidth]{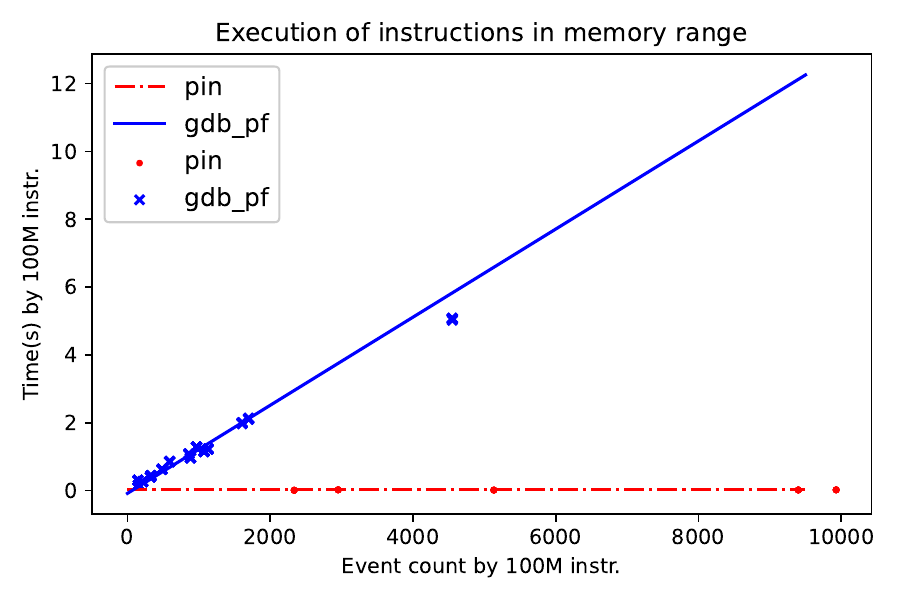}
\caption{Time required to execute 100 million instructions by number of
instruction executions in a memory range. Trend lines
represent the linear regression of the plotted points.}
\label{fig:instructions_in_page_single}
\end{figure}

Figure~\ref{fig:instructions_in_page_single} shows the execution time for every
100M instructions for programs with a different number of instrumentation events
in a memory range (i.e., a memory page) per 100M instructions. We observe
that while the execution time for page-fault-based instrumentation in GDB grows linearly
with the number of instrumented events, Pin remains almost constant. In this
case, Pin decides which instructions ought to be instrumented at translation
time, adding very little overhead since it only inserts 
callbacks at instructions that are part of the target page.

Regarding the overhead imposed by GDB single-stepping when instrumenting the
execution of all instructions (shown in Table~\ref{tbl:single_process} in
Appendix~\ref{appendix:res}), we can notice that execution time is extremely
high compared to Pin. In fact, as aforementioned, we had to limit GDB to 20M
instrumented events. If we compare the normalized instrumentation time (total
time minus the time required to execute the program with no instrumentation) for
20M instrumented events, GDB single-stepping is more than 100,000 times slower
than Pin at instruction tracing. Similar patterns can be noticed for
page-fault-based GDB. In that case, we had to limit GDB to track up to 200,000
events.

\begin{figure}[h]
\centering
\includegraphics[width=0.48\textwidth]{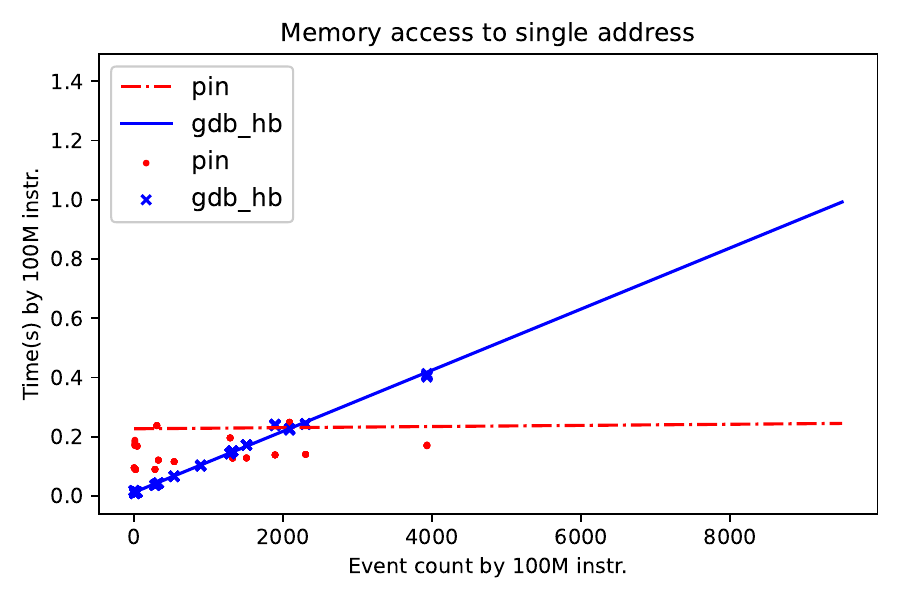}
\caption{Execution time for 100 million instructions by number of access events
per 100 million instructions. Trend lines represent the linear regression of
the plotted points.}
\label{fig:address_accesses_single}
\end{figure}

As detailed in previous sections, Pin has to check memory operations
on every instruction at run-time to decide if it affects a target memory range
or a memory address.

Figure~\ref{fig:address_accesses_single} shows the time required to execute 100M
instructions for our benchmark programs that trigger a different number of
address access events per 100M instructions. Even if hardware breakpoints are 
very efficient to instrument accesses to specific memory addresses, and
Pin needs to instrument every instruction, its overhead is
compensated as soon as the number of events triggered is higher than 2,000 every
100M instructions. This case is particularly interesting because the
\textit{cutting point} after which Pin performs better than
debugger-technique-based approaches is higher than in the cases shown previously
(e.g., single instrumented instruction execution).

\begin{figure}[h]
\centering
\includegraphics[width=0.48\textwidth]{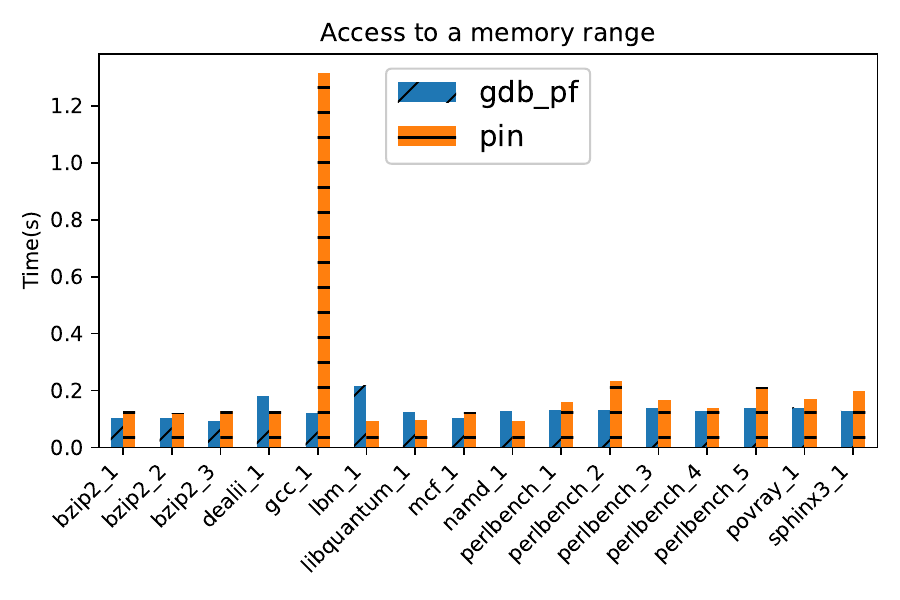}
\caption{Overhead of process-level DBI systems for memory range read/write.}
\label{fig:page_access_single}
\end{figure}

Figure~\ref{fig:page_access_single} shows the time required by GDB to instrument
100 page accesses using page faults for the different benchmark programs, and
the time it takes Pin to check the target address for every memory operation in
100M instructions. This overhead in Pin depends not only on the
number of instructions, but also on the type of operations and number of memory
accesses of these operations (and thus there is some variability in
Figure~\ref{fig:page_access_single}). However, we can see that Pin
significantly outperforms GDB except for very specific set-ups where memory
accesses are very frequent and the target memory page or target memory address
will be accessed extremely infrequently. This can be noticed in
Table~\ref{tbl:single_process} in Appendix~\ref{appendix:res}.

While using page faults to monitor accesses to a single memory
address (which is the same as that of accesses to a memory range) is
significantly slower than Pin and hardware breakpoint based approaches, this
difference is considerably smaller for the \texttt{lbm} benchmark, for which we
instrumented an address that is accessed very infrequently (i.e., 1,312 times).

\subsection{Whole-System Techniques Results}
\label{ss:results_whole_system}

Table~\ref{tbl:whole_system} in Appendix~\ref{appendix:res} shows the raw
results of our measurements on whole-system instrumentation techniques for each
primitive. In all cases, we report the average and standard deviation for each
iteration.

\begin{figure}[h]
\centering
\includegraphics[width=0.48\textwidth]{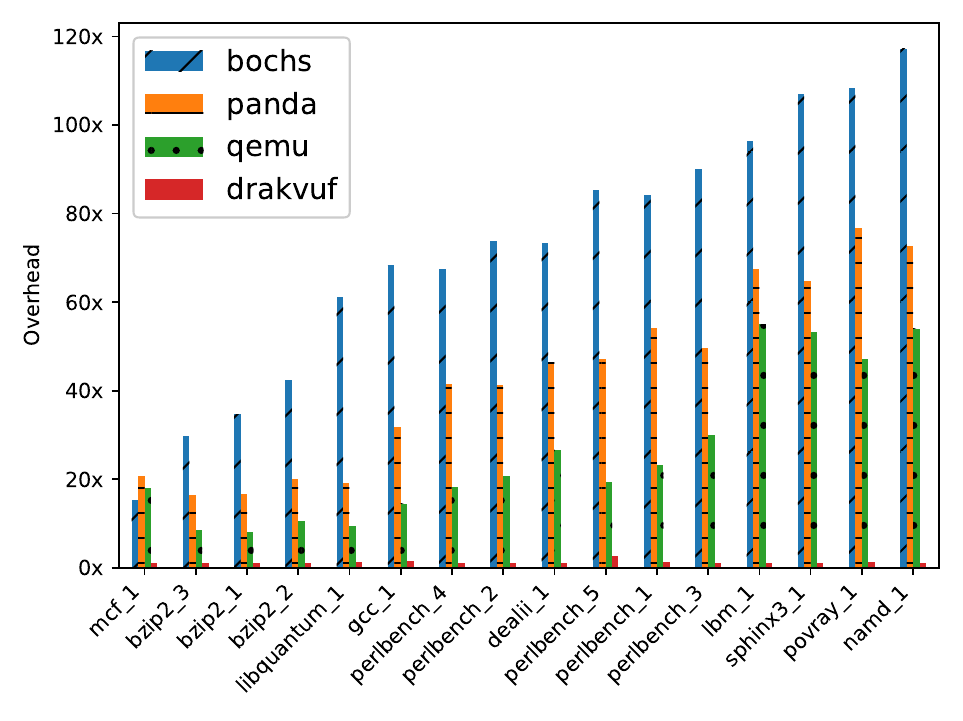}
\caption{Base overhead of each whole-system framework.}
\label{fig:overhead_whole}
\end{figure}

Figure~\ref{fig:overhead_whole} shows the base overhead of PANDA, Bochs and
Drakvuf with no instrumentation with respect to native execution. We also
include QEMU 2.9.1 (the version used in PANDA at the time of 
experimentation) as reference. We can notice that Drakvuf imposes negligible
and almost constant overhead, since hypervisor-based systems execute the guest
natively in the CPU. In the case of emulators, the
overhead is much higher and varies significantly, with the highest values found
for the floating-point intensive benchmarks.
Complex instruction set emulation is a non-trivial task, and its emulation can
inflict varying overheads depending on the complexity of the operations
performed by the program, the kind and frequency of memory accesses, and
optimizations. Overall, Bochs has higher overhead than PANDA and 
ranges from 20 to 120 times the native execution time. It is worth to note that
the \texttt{mcf} benchmark has the highest number of cache misses, and intensive pointer
and integer arithmetic.

\begin{figure}[h]
\centering
\includegraphics[width=0.48\textwidth]{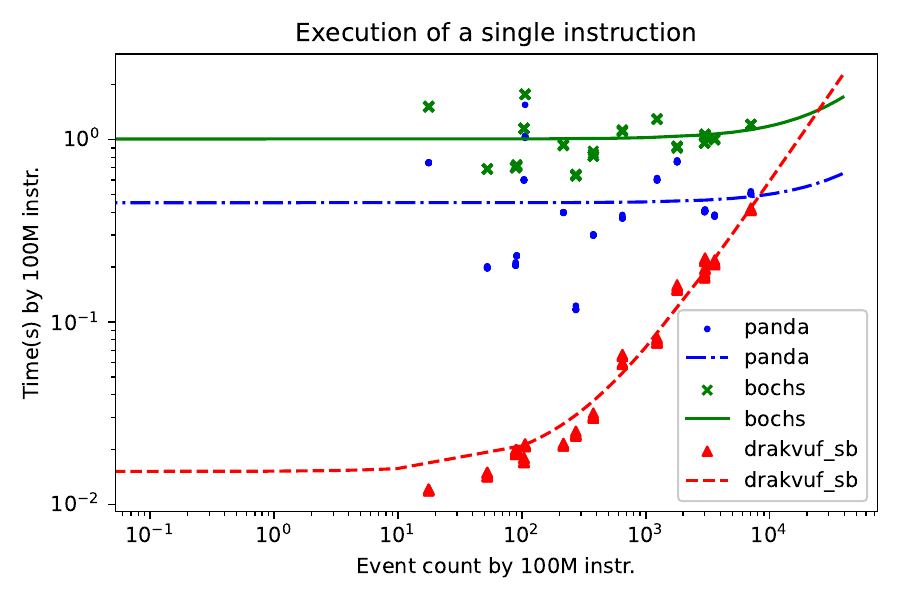}
\caption{Execution time for 100M instructions by number of instruction execution
events per 100M, in log scale. Trend lines represent the linear regression
of the plotted points.}
\label{fig:one_instruction_whole}
\end{figure}

Figure~\ref{fig:one_instruction_whole} shows the time required to execute 100M
instructions at different instrumentation frequencies for the single instruction
instrumentation experiment. The results are plotted in logarithmic scale. We 
notice a particular point in which Drakvuf becomes slower than PANDA and Bochs:
when the number of events per 100M instruction grows to 100,000. Similarly to
process-level instrumentation, the overhead of context switches --- and in this
case, VM exit operations --- has a non-negligible cost. This cost can compensate
emulation overhead after a certain point. There is also
more variability in execution times for Bochs and Panda due to the
different base overhead imposed by different programs (shown in
Figure~\ref{fig:overhead_whole}).

Figure~\ref{fig:inscount_whole}, Figure~\ref{fig:address_accesses_whole}, and
Figure~\ref{fig:every_memory_rw_whole} in Appendix~\ref{appendix:res} show the
execution times for the following primitives: execution of every
instruction, memory accesses to an address, and accesses to every memory
address respectively. In all cases, we observe similar performances for
Bochs and PANDA, which emulate memory in software. The superior overhead imposed
by Bochs in a non-instrumented execution is often compensated, but the results
depend on the type of program being instrumented. We can claim that results for
both tools in their current state suggest that the techniques at
instruction-level granularity are comparable, and that their relative overhead
is subject to specific implementation details and optimizations.

\begin{figure}[h]
\centering
\includegraphics[width=0.48\textwidth]{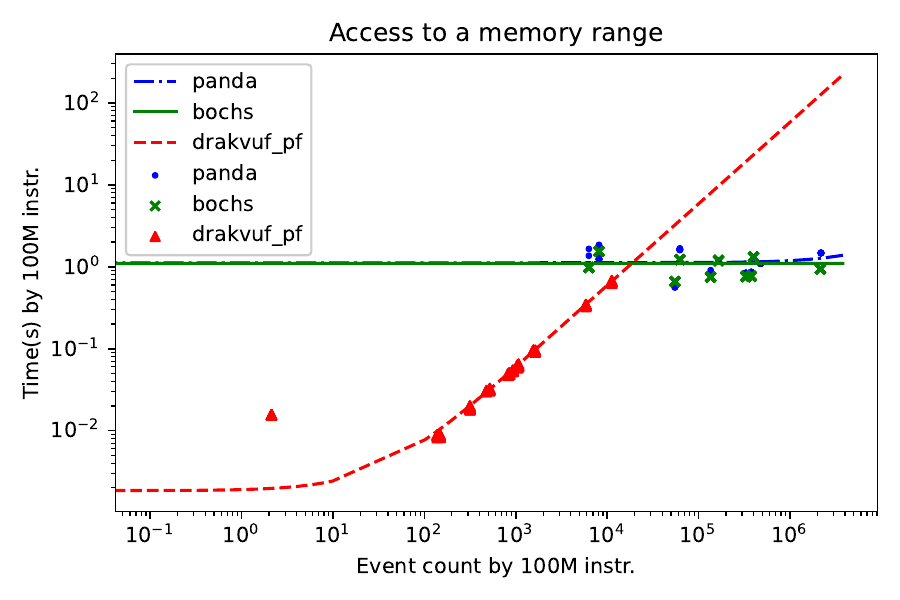}
\caption{Execution time for 100M instructions by number of memory access events per
100M instructions, in log scale. Trend lines represent the linear regression
of the plotted points.}
\label{fig:page_accesses_whole}
\end{figure}

\begin{figure}[h]
\centering
\includegraphics[width=0.48\textwidth]{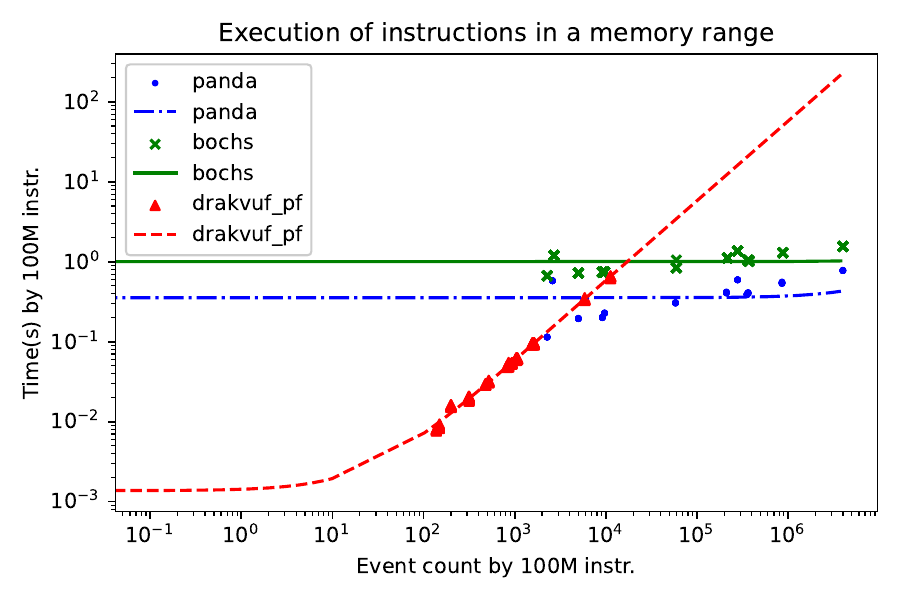}
\caption{Execution time for 100M instructions by number of instruction execution
events per 100M, in log scale. Trend lines represent the linear regression
of the plotted points.}
\label{fig:instructions_in_page_whole}
\end{figure}

Figure~\ref{fig:page_accesses_whole} and
Figure~\ref{fig:instructions_in_page_whole} show the normalized execution time
(time per 100M executed program instructions) for different instrumentation
event counts for the following experiments respectively: memory accesses to a
memory range (i.e., a memory page), and instruction execution in a given memory
range.  While increased execution time remains constant for most programs in
Drakvuf, the variability persists in Bochs and PANDA as in previous
cases. The axes are represented in logarithmic scale to improve
their legibility. As expected, there is a \textit{cutting point} beyond which
the execution overhead of Bochs and PANDA is compensated due to the penalty
imposed by the slower VM switches needed by Drakvuf.

\section{Insights from the Evaluation}
\label{sec:insights}

In some experiments, the overhead was so high that they would have required an
unaffordable and impractical amount of time to complete. In these cases, we
restricted the number of events to monitor to 20M in the case of
single-stepping, and 200K in the case of page faults to scale down execution
time. To present our results, we have normalized the overhead as needed in order
to compare approaches under the same conditions. We believe that these limits
are high enough to have statistically representative data. Finally, we do not
cover specific instrumentation use cases, but the actual primitives that those
use cases may rely on.

\subsection{General Insights}

\noindent \textbf{Instrumentation technique capabilities.} The fact that a given
instrumentation technique is able to instrument a specific primitive does not
mean that it is the most appropriate approach for that task. One clear example is
instrumentation through page faults. We have observed numerous times in our
experiments that page-fault-based instrumentation has a prohibitive overhead,
orders of magnitude higher than dynamic binary rewriting and instruction by
instruction monitoring. Our results show that there is not a clear winner for
every situation. However, it is fundamental to ascertain that the capabilities
of a given instrumentation technique match the specific requirements of the
problem at hand. In addition, factors to consider include the number of times
that a monitored event is expected to be triggered, and program characteristics
that may affect the performance of techniques and their implementations. These
characteristics include the amount and type of memory accesses, instruction
footprint, execution time, and floating-point operation intensity, as observed
during our experimentation.

\vspace*{0.3em}

\noindent \textbf{Emulation overhead variability.} Emulation and other binary
rewriting based approaches often need to decompose complex instructions such as
the ones in the Intel architecture (CISC) into smaller operations. Hardware
processors benefit from co-processors and hardware optimizations to execute
these instructions in a fast way, but emulators may introduce varying overheads
depending on the types of instructions. We have noticed this pattern in the
results for Pin, Bochs, and PANDA. For this reason, it is not only important to
consider the kind of instrumentation or the number of events that will be
triggered, but also the nature of the program and its base emulation overhead.

\vspace*{0.3em}

\noindent \textbf{Emulation overhead vs instrumentation overhead.} As mentioned
before, emulators and binary rewriting introduce a non-negligible overhead with
respect to native execution or hypervisor-based virtual machines. However, these
systems often run their instrumentation and callbacks in the same context and
address space as the emulated program. This saves numerous context switches and
interrupts, and therefore avoids wasting millions of cycles to deal with these
events. Furthermore, we have also observed that there is an expected increase in
baseline overhead for PANDA as compared to its base emulator QEMU. This overhead
difference is on average 1.8 times in our experiments, and can be attributed to
the infrastructure that enables instrumentation, execution of analysis
callbacks, and additional analyses via plugins. While emulators like Bochs or
PANDA are well known for their high overhead, we have noted that this negative
effect is soon compensated when the number of events to instrument is high.

\subsection{Technique-Specific Insights}

\noindent \textbf{JIT-based dynamic binary translation.} By design, the JIT-DBT
technique has a high upfront cost but it is very efficient once running. It
incurs a significant baseline slowdown (i.e., Pin up to 12 times the native
execution with no instrumentation), but per-instrumentation-event overhead is
low. As the number of instrumented events grows, JIT-DBT scales much better than
trap-based methods. For instance, Pin overtakes the breakpoint-based approach in
GDB after around 100 instrumented events per 100M instructions. Therefore,
JIT-DBT is more appropriate to monitor a high number of events (e.g., for heavy
code coverage or tracing), or when very low per-event overhead is critical,
since it handles large-scale instrumentation efficiently, not needing expensive
context switches. Conversely it is less suitable for very light workloads or
when minimal startup overhead is needed, since its initialization and analysis
costs dominate in short runs (e.g., trivial tasks or a few probes).

\vspace*{0.3em}

\noindent \textbf{Dynamic probe injection.} DPI has very low baseline overhead
close to native execution (similar to debugging-based DBI) and no per-event
context switches. In practice, DynInst easily outperforms Pin and GDB for sparse
instrumentation: in tests, as soon as a few events are instrumented, DPI
exhibits superior performance to JIT-DBT and trap-based approaches. Dynamic
probe injection works especially well for efficient instrumentation of a known
set of code locations with minimal overhead. It excels at instrumenting
well-defined code regions such as function entries where it can insert code
detours ahead of time. It provides near-native speed for instrumentation,
avoiding trap overhead from context switches and JIT costs. However, when the
number of individual instructions to be monitored is high, managing and
maintaining that many probes becomes impractical, since DPI requires a
trampoline at every instrumented location.

\vspace*{0.3em}

\noindent \textbf{Software and hardware breakpoints.} Similar to DPI,
breakpoint-based instrumentation has essentially zero cost when idle, since code
runs natively. Each breakpoint hit incurs a trap and full context switch into
the debugger however, so performance degrades rapidly as events increase. In
experiments, breakpoint overhead grows linearly with event count and quickly
exceeds JIT-DBT. In particular, by approximately 100 breakpoints per 100M
instructions, Pin performs better than GDB. For memory access instrumentation,
hardware breakpoints have shown to be very efficient for a single address: tests
showed that the breakpoint-based approach was faster than JIT-DBT until about
2,000 hits per 100M instructions. Results suggest that breakpoints perform
better for sparse events or isolated instrumentation points (e.g., a small
number of code locations or memory addresses). On the contrary, breakpoints are
less convenient when the instrumentation events are frequent or high-rate, since
the trap and context switch overhead will dominate.

\vspace*{0.3em}

\noindent \textbf{Single-stepping.} Traps on every instruction yield the highest
overhead. In practice, GDB single-stepping was orders of magnitude slower than
JIT-DBT in Pin for tens of millions of steps. For almost any realistic workload,
this method should be impractically slow, making other techniques such as
JIT-DBT or an interpretation-based approach a more efficient and flexible option.
Single-stepping should be considered as a supporting technique to other
instrumentation methods, as a last resort when there are limited alternatives,
for tiny workloads, or when formal correctness of each instruction observation
is required.

\vspace*{0.3em}

\noindent \textbf{Page fault traps.} Memory protection traps are similar in
concept to breakpoints. Baseline overhead is low, but each page fault trap
incurs context switching overhead. In the experimentation, page-fault-based
instrumentation overhead grew with the number of page hits, whereas JIT-DBT
remained almost flat by inserting checks at translation time for
instruction-related tests and at execution time for memory-related tests.
Page-fault-based instrumentation works best for a reduced number of page-level
events or when hardware breakpoints are not available (e.g., instrumenting
ranges of memory). However, it is less adequate if events are frequent or
fine-grained, since its cost scales with event rate as with breakpoints. For
example, Pin imposed almost no extra cost per page hit, while the page fault
approach in GDB slowed down proportionally with the number of faults.

\vspace*{0.3em}

\noindent \textbf{CPU interpretation.} A CPU interpreter or full-system emulator
pays a huge penalty in exchange for being almost equally flexible as JIT-DBT.
For instance, Bochs was on the order of 20 to 120 times slower than native
execution with no instrumentation. CPU interpretation is best suited when strict
execution control is required (e.g., full-system analysis, instruction-level
verification) and performance is less relevant, the frequency of instrumentation
events is significant, or when hardware support is not available. Since CPU
interpretation and JIT-DBT follow a similar approach for memory access
instrumentation, they exhibit analogous performance in these applications. For a
lower number of instrumentation events, trap-based methods perform better, as
well as JIT-DBT for instruction-related instrumentation.

\section{Discussion}
\label{sec:dis}

Several papers have covered DBI in the past, focusing on the transparency of
specific approaches in adversarial environments rather than their foundations,
convenience, capabilities and efficiency. 

In this paper, we made an effort to isolate instrumentation techniques
from the tools that implement them, but the results show that implementation
details have certain influence on the performance. For instance, tools such as
GDB inevitably add some overhead from having to notify the debugger on each
event and so on. This means that some of the \textit{cutting points} that we
show must be interpreted as guidelines and not as absolute numbers. However,
even though there may be some variability in the results due to specific tool
and environment features, we believe that the conclusions presented would not
change significantly.

\vspace*{0.3em}

\noindent \textbf{Hardware support in DBI.} Hardware features (e.g., shadow page
tables, EPT, debug registers, PMU) are strongly used and convenient in DBI, and
usually guide the development of different implementations of existing
instrumentation techniques and new DBI frameworks. For instance, the PowerPC
architecture supports ranged breakpoints (apart from individual addresses).
However, these are not supported on x86, which impedes existing and new DBI
tools from taking advantage of them to efficiently implement range primitives.
Similarly, some Intel processors include a feature called Memory Protection Keys
(MPK) that enable fast memory protection management at page-level~\cite{mpk}. We
have observed that page-fault-based instrumentation is not practical in most
scenarios due to its performance, but features such as MPK enable more efficient
and functional DBI (i.e., because it does not require modification of the page
tables). At the hypervisor-level, most frameworks rely on VMEXIT to implement
all instrumentation logic. Nevertheless, VMFUNC is another Intel primitive that
allows to change EPT page tables in a VM without exiting into the hypervisor.
This enables and guides the implementation of more efficient hypervisor-based
DBI~\cite{shi2018shadowmonitor,hong2021novel}. Instrumentation techniques evolve
to leverage these advanced features to provide higher flexibility, lower
overhead, and more robust system-wide analysis.

\vspace*{0.3em}

\noindent \textbf{Capabilities, performance and transparency trade-offs.} The
main properties that define a DBI framework are the instrumentation
expressiveness, transparency, and performance. Existing approaches make
trade-offs in some of these properties in exchange for enhancements in others.
For instance, they often have to choose between low overhead and high
transparency or fine-grained capabilities. It is worth mentioning that JIT-DBT
is the only technique that implements coarse-grained primitives (i.e., at block
level) by its nature, convenient for tasks such as coverage retrieval in
fuzzing. There are further aspects that also play a role in the process of
developing or choosing a tool for a specific task, such as platform and
architecture compatibility, and usability. Users may favour an implementation
over another based on usability considerations regardless of the overhead.
However, most of these aspects are generally implementation dependent and not
related to the instrumentation techniques.

\vspace*{0.3em}

\noindent \textbf{Maturation of the field.} Process-level DBI has found
widespread use in profiling, debugging, and security research due to its
practicality (e.g., working in user-space, having fewer hardware
dependencies). Historically, whole-system DBI was driven by the needs of malware
analysis and reverse engineering in controlled environments. With some academic
works focusing on specific case studies rather than developing broadly
applicable frameworks, fewer general-purpose whole-system tools have seen
ongoing development. The apparent slowdown in fundamental DBI research and the
uneven maintenance of process-level versus whole-system tools reflect both the
maturation of the field and the shifting priorities of the community. Many of
the core techniques in DBI have been extensively studied and have reached a
degree of maturity, incremental improvements are still happening, but they may
be more focused on integration with other advanced analysis techniques, on
improving resilience towards adversaries, or on specialized use cases rather
than on fundamental DBI.

\section{Related Work}
\label{sec:relatedwork}

Prior work on DBI has focused on DBI in specific application domains,
transparency properties of DBI systems, or specific instrumentation techniques
(e.g., measuring the internal performance of Pin~\cite{uh2006analyzing}).
Moreover, other works related to ours include papers that analyze an individual
building block of DBI or features used by instrumentation techniques.

\textbf{DBI in Malware Research.} Malware research is one of the most popular
application domains of DBI, especially for whole-system approaches
\cite{egele2012survey,lengyel2014drakvuf}. Several works have
been devoted to analyze over a decade of malware analysis evasion techniques
that hinder DBI and their
countermeasures~\cite{shi2014cardinal,polino2017measuring,bulazel2017survey,afianian2018malware,d2022evaluating},
while others are more focused on specific tools~\cite{kirsch2018pwin}. From a
different perspective, previous work has also focused on the definition and
fulfillment of transparency requirements for DBI
systems~\cite{dinaburg2008ether,ning2017ninja,d2019sok}. These works analyze
specific aspects of DBI and different usability circumstances. Our work is
different in nature since we analyze DBI as a whole and fundamental aspects such
as instrumentation capabilities.

\textbf{DBI Elements and Analysis.} Jain et al.~\cite{jain2014sok} proposed an
analysis of introspection capabilities and approaches, describing
the different VMI designs, the evolution of the semantic gap problem and
proposed solutions, and the various attacks against VMI, as well as defenses,
and trust assumptions. Since introspection is a building block of DBI, we draw
on this systematization of knowledge to express introspection capabilities of
whole-system approaches.

A work close to ours~\cite{d2019sok}, describes the inner workings and
primitives of DBI frameworks, their usage in security research, and discusses
the transparency of such systems along with evasion and escape techniques,
proposing some mitigations. In contrast to our work, they analyze JIT-based
user-space DBI frameworks, and focus on the transparency issues of DBI, whereas
we take a broader approach by analyzing every technique, and focusing on
instrumentation capabilities and performance.

\section{Conclusion}
\label{sec:conc}

DBI systems are largely applied in a broad range of areas, especially in
security research. Numerous approaches have been
proposed, implementing different techniques and designs with their particular
advantages and limitations.

With this paper we aimed to shed light into the heterogeneous ecosystem of DBI.
Our work is the first to lay out the building blocks and
foundations of DBI, and to analyze existing instrumentation techniques.
We evaluated their performance based on practical implementations
for both process-level and whole-system, and showed
that there is no one-fits-all approach.
We believe that this paper will bring a better understanding
of the inner workings of DBI, and guide future research to adapt or develop new
instrumentation techniques and frameworks.

\section*{Acknowledgments}
\label{sec:acknowledgement}
This work is partially supported by the Basque Government under a pre-doctoral
grant given to Oscar Llorente-Vazquez.

\bibliographystyle{plain}
\bibliography{paper}

\begin{thebibliography}{10}

\bibitem{mpk}
{Memory Protection Keys - The Linux kernel Documentation}.
\newblock https://www.kernel.org/doc/html/latest/core-api/protection-keys.html,
  2022.

\bibitem{afianian2018malware}
Amir Afianian, Salman Niksefat, Babak Sadeghiyan, and David Baptiste.
\newblock {Malware Dynamic Analysis Evasion Techniques: A Survey}.
\newblock {\em {ACM Computing Surveys (CSUR)}}, 52(6):1--28, 2019.

\bibitem{attariyan2012x}
Mona Attariyan, Michael Chow, and Jason Flinn.
\newblock {X-ray: Automating Root-Cause Diagnosis of Performance Anomalies in
  Production Software}.
\newblock In {\em {Proceedings of the USENIX Symposium on Operating Systems
  Design and Implementation (OSDI)}}, 2012.

\bibitem{bellard2005qemu}
Fabrice Bellard.
\newblock {QEMU, a Fast and Portable Dynamic Translator}.
\newblock In {\em {USENIX Annual Technical Conference, FREENIX Track}},
  volume~41, 2005.

\bibitem{bernat2011anywhere}
Andrew~R Bernat and Barton~P Miller.
\newblock {Anywhere, Any-Time Binary Instrumentation}.
\newblock In {\em {Proceedings of the ACM SIGPLAN-SIGSOFT Workshop on Program
  Analysis for Software Tools}}, 2011.

\bibitem{bruening2004rio}
Derek Bruening.
\newblock {\em {Efficient, Transparent, and Comprehensive Runtime Code
  Manipulation}}.
\newblock PhD thesis, {Massachusetts Institute of Technology}, 2004.

\bibitem{bulazel2017survey}
Alexei Bulazel and B{\"u}lent Yener.
\newblock {A Survey on Automated Dynamic Malware Analysis Evasion and
  Counter-Evasion: PC, Mobile, and Web}.
\newblock In {\em {Proceedings of the Reversing and Offensive-oriented Trends
  Symposium}}, 2017.

\bibitem{chipounov2011s2e}
Vitaly Chipounov, Volodymyr Kuznetsov, and George Candea.
\newblock {S2E: A Platform for In-Vivo Multi-Path Analysis of Software
  Systems}.
\newblock In {\em {Proceedings of the International Conference on Architectural
  Support for Programming Languages and Operating Systems (ASPLOS)}}, 2011.

\bibitem{clements2020halucinator}
Abraham~A Clements, Eric Gustafson, Tobias Scharnowski, Paul Grosen, David
  Fritz, Christopher Kruegel, Giovanni Vigna, Saurabh Bagchi, and Mathias
  Payer.
\newblock {HALucinator: Firmware Re-hosting Through Abstraction Layer
  Emulation}.
\newblock In {\em {Proceedings of the USENIX Security Symposium}}, 2020.

\bibitem{rekall}
Michael Cohen.
\newblock {The Rekall Forensic and Incident Response Framework}.
\newblock https://github.com/google/rekall, 2022.

\bibitem{d2019sok}
Daniele~Cono D'Elia, Emilio Coppa, Simone Nicchi, Federico Palmaro, and Lorenzo
  Cavallaro.
\newblock {SoK: Using Dynamic Binary Instrumentation for Security (And How You
  May Get Caught Red Handed)}.
\newblock In {\em {Proceedings of the ACM on Asia Conference on Computer and
  Communications Security (AsiaCCS)}}, 2019.

\bibitem{deng2013spider}
Zhui Deng, Xiangyu Zhang, and Dongyan Xu.
\newblock {Spider: Stealthy Binary Program Instrumentation and Debugging via
  Hardware Virtualization}.
\newblock In {\em {Proceedings of the Annual Computer Security Applications
  Conference}}, 2013.

\bibitem{dinaburg2008ether}
Artem Dinaburg, Paul Royal, Monirul Sharif, and Wenke Lee.
\newblock {Ether: Malware Analysis via Hardware Virtualization Extensions}.
\newblock In {\em {Proceedings of the ACM Conference on Computer and
  Communications Security}}, 2008.

\bibitem{dolan2015panda}
Brendan Dolan-Gavitt, Josh Hodosh, Patrick Hulin, Tim Leek, and Ryan Whelan.
\newblock {Repeatable Reverse Engineering with PANDA}.
\newblock In {\em {Proceedings of the Program Protection and Reverse
  Engineering Workshop}}, 2015.

\bibitem{dolan2011virtuoso}
Brendan Dolan-Gavitt, Tim Leek, Michael Zhivich, Jonathon Giffin, and Wenke
  Lee.
\newblock {Virtuoso: Narrowing the Semantic Gap in Virtual Machine
  Introspection}.
\newblock In {\em {Proceedings of the IEEE Symposium on Security and Privacy}},
  2011.

\bibitem{d2022evaluating}
Daniele~Cono D’Elia, Lorenzo Invidia, Federico Palmaro, and Leonardo
  Querzoni.
\newblock {Evaluating Dynamic Binary Instrumentation Systems for Conspicuous
  Features and Artifacts}.
\newblock {\em Digital Threats: Research and Practice (DTRAP)}, 3(2):1--13,
  2022.

\bibitem{egele2012survey}
Manuel Egele, Theodoor Scholte, Engin Kirda, and Christopher Kruegel.
\newblock {A Survey on Automated Dynamic Malware-Analysis Techniques and
  Tools}.
\newblock {\em {ACM Computing Surveys (CSUR)}}, 44(2):6, 2012.

\bibitem{filho2022evasion}
Ailton~Santos Filho, Ricardo~J Rodr{\'\i}guez, and Eduardo~L Feitosa.
\newblock {Evasion and Countermeasures Techniques to Detect Dynamic Binary
  Instrumentation Frameworks}.
\newblock {\em Digital Threats: Research and Practice (DTRAP)}, 3(2):1--28,
  2022.

\bibitem{fu2012space}
Yangchun Fu and Zhiqiang Lin.
\newblock {Space Traveling across VM: Automatically Bridging the Semantic Gap
  in Virtual Machine Introspection via Online Kernel Data Redirection}.
\newblock In {\em {Proceedings of the IEEE Symposium on Security and Privacy}},
  2012.

\bibitem{gdb}
{GDB}.
\newblock {GDB: The GNU Project Debugger}.
\newblock https://www.gnu.org/software/gdb/, 2022.

\bibitem{tinyinst}
Google.
\newblock {TinyInst - A lightweight dynamic instrumentation library}.
\newblock https://github.com/googleprojectzero/TinyInst, 2022.

\bibitem{greathouse2012case}
Joseph~L Greathouse, Hongyi Xin, Yixin Luo, and Todd Austin.
\newblock {A Case for Unlimited Watchpoints}.
\newblock In {\em {Proceedings of the International Conference on Architectural
  Support for Programming Languages and Operating Systems (ASPLOS)}}, 2012.

\bibitem{henderson2014make}
Andrew Henderson, Aravind Prakash, Lok~Kwong Yan, Xunchao Hu, Xujiewen Wang,
  Rundong Zhou, and Heng Yin.
\newblock {Make It Work, Make It Right, Make It Fast: Building a
  Platform-Neutral Whole-System Dynamic Binary Analysis Platform}.
\newblock In {\em {Proceedings of the International Symposium on Software
  Testing and Analysis}}, 2014.

\bibitem{idapro}
{Hex-Rays}.
\newblock {IDA Pro}.
\newblock https://www.hex-rays.com/products/ida/, 2022.

\bibitem{hong2021novel}
Jiaqi Hong and Xuhua Ding.
\newblock {A Novel Dynamic Analysis Infrastructure to Instrument Untrusted
  Execution Flow Across User-kernel Spaces}.
\newblock In {\em {Proceedings of the IEEE Symposium on Security and Privacy}},
  2021.

\bibitem{jain2014sok}
Bhushan Jain, Mirza~Basim Baig, Dongli Zhang, Donald~E Porter, and Radu Sion.
\newblock {SoK: Introspections on Trust and the Semantic Gap}.
\newblock In {\em {Proceedings of the IEEE Symposium on Security and Privacy}},
  2014.

\bibitem{jurczyk2013identifying}
Mateusz Jurczyk and Gynvael Coldwind.
\newblock {Identifying and Exploiting Windows Kernel Race Conditions via Memory
  Access Patterns}.
\newblock 2013.

\bibitem{karvandi2022hyperdbg}
Mohammad~Sina Karvandi, MohammadHosein Gholamrezaei, Saleh Khalaj~Monfared,
  Soroush Meghdadizanjani, Behrooz Abbassi, Ali Amini, Reza Mortazavi, Saeid
  Gorgin, Dara Rahmati, and Michael Schwarz.
\newblock {HyperDbg: Reinventing Hardware-Assisted Debugging}.
\newblock In {\em {Proceedings of the ACM Conference on Computer and
  Communications Security}}, 2022.

\bibitem{ketterlin2012profiling}
Alain Ketterlin and Philippe Clauss.
\newblock Profiling data-dependence to assist parallelization: Framework,
  scope, and optimization.
\newblock In {\em {Proceedings of the Annual IEEE/ACM International Symposium
  on Microarchitecture (MICRO)}}, 2012.

\bibitem{kirsch2018pwin}
Julian Kirsch, Zhechko Zhechev, Bruno Bierbaumer, and Thomas Kittel.
\newblock {PwIN--Pwning Intel piN: Why DBI is Unsuitable for Security
  Applications}.
\newblock In {\em {European Symposium on Research in Computer Security
  (ESORICS)}}, 2018.

\bibitem{lawton1996bochs}
Kevin~P Lawton.
\newblock {Bochs: A Portable PC Emulator for Unix/X}.
\newblock {\em Linux Journal}, 1996(29es):7, 1996.

\bibitem{lengyel2014drakvuf}
Tamas~K Lengyel, Steve Maresca, Bryan~D Payne, George~D Webster, Sebastian
  Vogl, and Aggelos Kiayias.
\newblock {Scalability, Fidelity and Stealth in the DRAKVUF Dynamic Malware
  Analysis System}.
\newblock In {\em {Proceedings of the Annual Computer Security Applications
  Conference}}, 2014.

\bibitem{luk2005pin}
Chi-Keung Luk, Robert Cohn, Robert Muth, Harish Patil, Artur Klauser, Geoff
  Lowney, Steven Wallace, Vijay~Janapa Reddi, and Kim Hazelwood.
\newblock {Pin: Building Customized Program Analysis Tools with Dynamic
  Instrumentation}.
\newblock In {\em {Proceedings of the ACM SIGPLAN Conference on Programming
  Language Design and Implementation}}, 2005.

\bibitem{miramirkhani2017spotless}
Najmeh Miramirkhani, Mahathi~Priya Appini, Nick Nikiforakis, and Michalis
  Polychronakis.
\newblock {Spotless Sandboxes: Evading Malware Analysis Systems Using
  Wear-and-Tear Artifacts}.
\newblock In {\em {Proceedings of the IEEE Symposium on Security and Privacy}},
  2017.

\bibitem{nick2007valgrind}
Nicholas Nethercote and Julian Seward.
\newblock {Valgrind: A Framework for Heavyweight Dynamic Binary
  Instrumentation}.
\newblock In {\em {Proceedings of the ACM SIGPLAN Conference on Programming
  Language Design and Implementation}}, 2007.

\bibitem{nguyen2009mavmm}
Anh~M Nguyen, Nabil Schear, HeeDong Jung, Apeksha Godiyal, Samuel~T King, and
  Hai~D Nguyen.
\newblock {MAVMM: Lightweight and Purpose Built VMM for Malware Analysis}.
\newblock In {\em {Proceedings of the Annual Computer Security Applications
  Conference}}, 2009.

\bibitem{ning2017ninja}
Zhenyu Ning and Fengwei Zhang.
\newblock {Ninja: Towards Transparent Tracing and Debugging on ARM}.
\newblock In {\em {Proceedings of the USENIX Security Symposium}}, 2017.

\bibitem{pyemu}
Cody Pierce.
\newblock {PyEmu: x86 Emulator in Python}.
\newblock https://github.com/codypierce/pyemu, 2022.

\bibitem{polino2017measuring}
Mario Polino, Andrea Continella, Sebastiano Mariani, Stefano D’Alessio,
  Lorenzo Fontana, Fabio Gritti, and Stefano Zanero.
\newblock {Measuring and Defeating Anti-Instrumentation-Equipped Malware}.
\newblock In {\em {Proceedings of the International Conference on Detection of
  Intrusions and Malware, and Vulnerability Assessment (DIMVA)}}, 2017.

\bibitem{prakash2015vfguard}
Aravind Prakash, Xunchao Hu, and Heng Yin.
\newblock {vfGuard: Strict Protection for Virtual Function Calls in COTS C++
  Binaries}.
\newblock In {\em {Proceedings of the Network and Distributed System Security
  Symposium (NDSS)}}, 2015.

\bibitem{proskurin2018hiding}
Sergej Proskurin, Tamas Lengyel, Marius Momeu, Claudia Eckert, and Apostolis
  Zarras.
\newblock {Hiding in the Shadows: Empowering ARM for Stealthy Virtual Machine
  Introspection}.
\newblock In {\em {Proceedings of the Annual Computer Security Applications
  Conference}}, 2018.

\bibitem{pustogarov2020ex}
Ivan Pustogarov, Qian Wu, and David Lie.
\newblock {Ex-vivo dynamic analysis framework for Android device drivers}.
\newblock In {\em {Proceedings of the IEEE Symposium on Security and Privacy}},
  2020.

\bibitem{unicorn}
Nguyen~Anh Quynh and Dang~Hoang Vu.
\newblock {Unicorn CPU emulator framework}.
\newblock https://github.com/unicorn-engine/unicorn, 2022.

\bibitem{frida}
Ole Andre~Vadla Ravnas.
\newblock {Frida - A world-class dynamic instrumentation framework}.
\newblock https://www.frida.re/, 2022.

\bibitem{rvmi}
{rVMI}.
\newblock {rVMI - A New Paradigm For Full System Analysis}.
\newblock https://github.com/fireeye/rvmi, 2017.

\bibitem{saberi2014hybrid}
Alireza Saberi, Yangchun Fu, and Zhiqiang Lin.
\newblock {Hybrid-Bridge: Efficiently Bridging the Semantic Gap in Virtual
  Machine Introspection via Decoupled Execution and Training Memoization}.
\newblock In {\em {Proceedings of the Network and Distributed System Security
  Symposium (NDSS)}}, 2014.

\bibitem{shi2018shadowmonitor}
Bin Shi, Lei Cui, Bo~Li, Xudong Liu, Zhiyu Hao, and Haiying Shen.
\newblock {ShadowMonitor: An Effective in-VM Monitoring Framework with
  Hardware-enforced Isolation}.
\newblock In {\em {Proceedings of the International Symposium on Research in
  Attacks, Intrusions, and Defenses}}, 2018.

\bibitem{shi2014cardinal}
Hao Shi, Abdulla Alwabel, and Jelena Mirkovic.
\newblock {Cardinal Pill Testing of System Virtual Machines}.
\newblock In {\em {Proceedings of the USENIX Security Symposium}}, 2014.

\bibitem{ugarte2016rambo}
Xabier Ugarte-Pedrero, Davide Balzarotti, Igor Santos, and Pablo~G Bringas.
\newblock {RAMBO: Run-Time Packer Analysis with Multiple Branch Observation}.
\newblock In {\em {Proceedings of the International Conference on Detection of
  Intrusions and Malware, and Vulnerability Assessment}}, 2016.

\bibitem{pyrebox}
{Ugarte-Pedrero, Xabier}.
\newblock {PyREBox: Python scriptable Reverse Engineering Sandbox, a Virtual
  Machine instrumentation and inspection framework based on QEMU}.
\newblock https://github.com/Cisco-Talos/pyrebox, 2022.

\bibitem{uh2006analyzing}
Gang-Ryung Uh, Robert Cohn, Bharadwaj Yadavalli, Ramesh Peri, and Ravi
  Ayyagari.
\newblock {Analyzing Dynamic Binary Instrumentation Overhead}.
\newblock In {\em {Workshop on Binary Instrumentation and Applications (WBIA)
  at ASPLOS}}, 2006.

\bibitem{van2016tough}
Victor Van Der~Veen, Enes G{\"o}ktas, Moritz Contag, Andre Pawoloski, Xi~Chen,
  Sanjay Rawat, Herbert Bos, Thorsten Holz, Elias Athanasopoulos, and Cristiano
  Giuffrida.
\newblock {A Tough Call: Mitigating Advanced Code-Reuse Attacks at the Binary
  Level}.
\newblock In {\em {Proceedings of the IEEE Symposium on Security and Privacy}},
  2016.

\bibitem{vasudevan2005stealth}
Amit Vasudevan and Ramesh Yerraballi.
\newblock {Stealth Breakpoints}.
\newblock In {\em {Proceedings of the Annual Computer Security Applications
  Conference}}, 2005.

\bibitem{vasudevan2006cobra}
Amit Vasudevan and Ramesh Yerraballi.
\newblock {Cobra: Fine-grained Malware Analysis using Stealth
  Localized-executions}.
\newblock In {\em {Proceedings of the IEEE Symposium on Security and Privacy}},
  2006.

\bibitem{vogl2012using}
Sebastian Vogl and Claudia Eckert.
\newblock {Using Hardware Performance Events for Instruction-Level Monitoring
  on the x86 Architecture}.
\newblock In {\em {Proceedings of the European Workshop on Systems Security
  (EuroSec)}}, 2012.

\bibitem{wei2014quantifying}
Jiesheng Wei, Anna Thomas, Guanpeng Li, and Karthik Pattabiraman.
\newblock {Quantifying the Accuracy of High-Level Fault Injection Techniques
  for Hardware Faults}.
\newblock In {\em {Proceedings of the IEEE/IFIP International Conference on
  Dependable Systems and Networks (DSN)}}, 2014.

\bibitem{xiao2017stacco}
Yuan Xiao, Mengyuan Li, Sanchuan Chen, and Yinqian Zhang.
\newblock {STACCO: Differentially Analyzing Side-Channel Traces for Detecting
  SSL/TLS Vulnerabilities in Secure Enclaves}.
\newblock In {\em {Proceedings of the ACM Conference on Computer and
  Communications Security (CCS)}}, 2017.

\bibitem{yan2012v2e}
Lok-Kwong Yan, Manjukumar Jayachandra, Mu~Zhang, and Heng Yin.
\newblock {V2E: Combining Hardware Virtualization and Software Emulation for
  Transparent and Extensible Malware Analysis}.
\newblock In {\em {Proceedings of the ACM SIGPLAN/SIGOPS International
  Conference on Virtual Execution Environments}}, 2012.

\bibitem{yin2010temu}
Heng Yin and Dawn Song.
\newblock {TEMU: Binary Code Analysis via Whole-System Layered Annotative
  Execution}.
\newblock {\em EECS Department, University of California, Berkeley, Tech. Rep.
  UCB/EECS-2010-3}, 2010.

\bibitem{zeng2015pemu}
Junyuan Zeng, Yangchun Fu, and Zhiqiang Lin.
\newblock {PEMU: A Pin Highly Compatible Out-of-VM Dynamic Binary
  Instrumentation Framework}.
\newblock In {\em Proceedings of the 11th ACM SIGPLAN/SIGOPS International
  Conference on Virtual Execution Environments}, 2015.

\bibitem{zhang2015malt}
Fengwei Zhang, Kevin Leach, Angelos Stavrou, Haining Wang, and Kun Sun.
\newblock {Using Hardware Features for Increased Debugging Transparency}.
\newblock In {\em {Proceedings of the IEEE Symposium on Security and Privacy}},
  2015.

\bibitem{zhao2017seeing}
Siqi Zhao, Xuhua Ding, Wen Xu, and Dawu Gu.
\newblock {Seeing Through The Same Lens: Introspecting Guest Address Space At
  Native Speed}.
\newblock In {\em {Proceedings of the USENIX Security Symposium}}, 2017.

\end{thebibliography}

\appendix
\section*{Appendix}
\section{Complementary Evaluation Results}
\label{appendix:res}

We measured the run-time performance overhead of the existing instrumentation
techniques within DBI. To this end, we selected a set of existing DBI tools that
implement these techniques, seeking to encompass all techniques that are
currently implemented in existing frameworks for both process-level and
whole-system DBI. The full list of used implementations along with specific
details is presented in \textsection\ref{sec:eval}.

In our experiments, we prepared a set of elemental instrumentation tests that
are directly correlated to instrumentation primitives to better evaluate the
fundamentals of the instrumentation techniques. Our launcher program is the
responsible for managing and automating experiments execution, while also
collecting a variety of data from the executions (e.g., cpu-time and wall-clock
time).

Next, we present the complete results for the corpus of C/C++ benchmarks in the
SPEC CPU2006 suite that we used for the measurement. Out of the suite, we used a
total of 16 benchmarks from 10 unique binaries, namely: \textit{perlbench\_1},
\textit{perlbench\_2}, \textit{perlbench\_3}, \textit{perlbench\_4},
\textit{perlbench\_5}, \textit{bzip2\_1}, \textit{bzip2\_2}, \textit{bzip2\_3},
\textit{gcc}, \textit{mcf}, \textit{libquantum}, \textit{namd}, \textit{dealii},
\textit{povray}, \textit{lbm}, and \textit{sphinx3}, as described in
\textsection\ref{sec:eval}.

\begin{figure}[htp]
\centering
\includegraphics[width=0.48\textwidth]{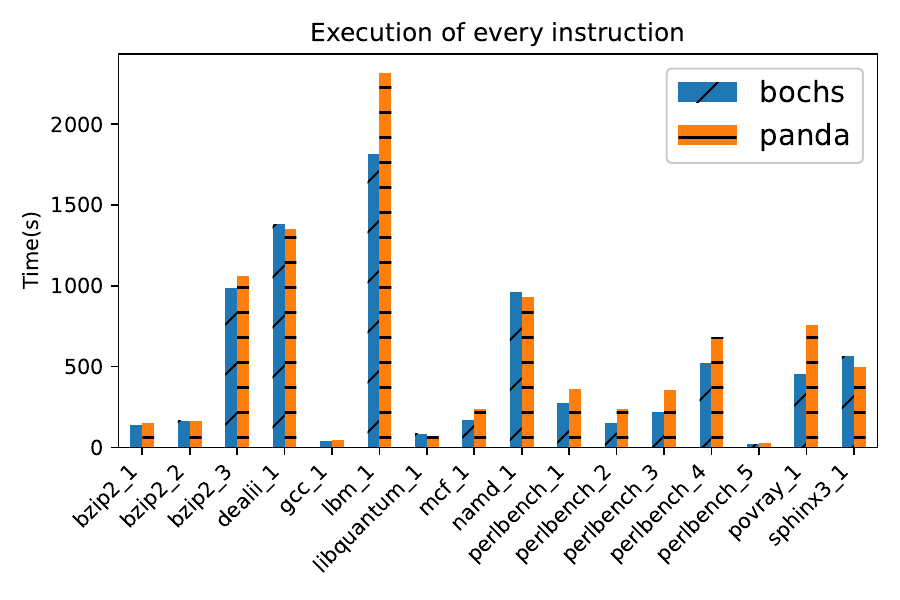}
\caption{Execution time for instrumenting every instruction in each benchmark
program for whole-system tools.}
\label{fig:inscount_whole}
\end{figure}

\begin{figure}[htp]
\centering
\includegraphics[width=0.48\textwidth]{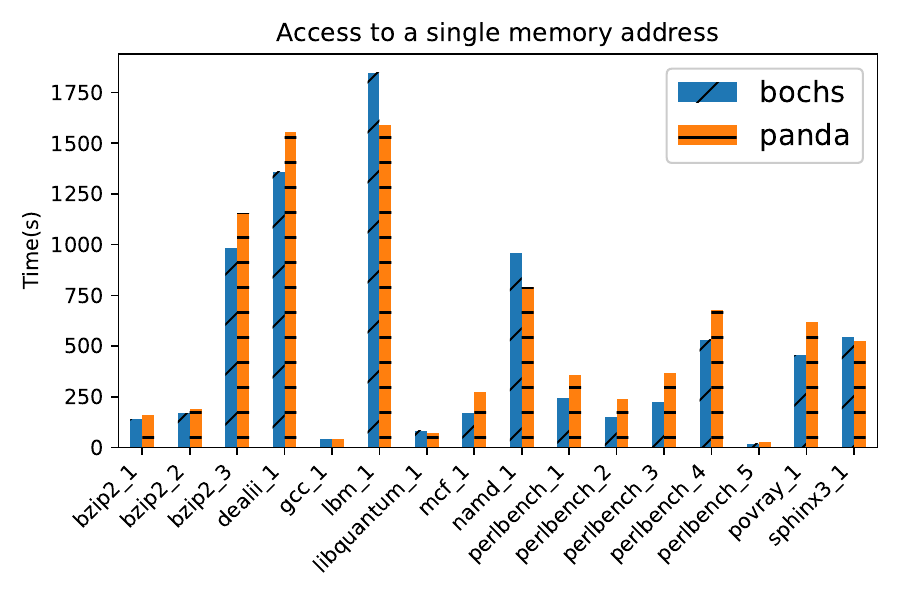}
\caption{Execution time for instrumented accesses to a memory address in
benchmark programs for whole-system.}
\label{fig:address_accesses_whole}
\end{figure}

\begin{figure}[htp]
\centering
\includegraphics[width=0.48\textwidth]{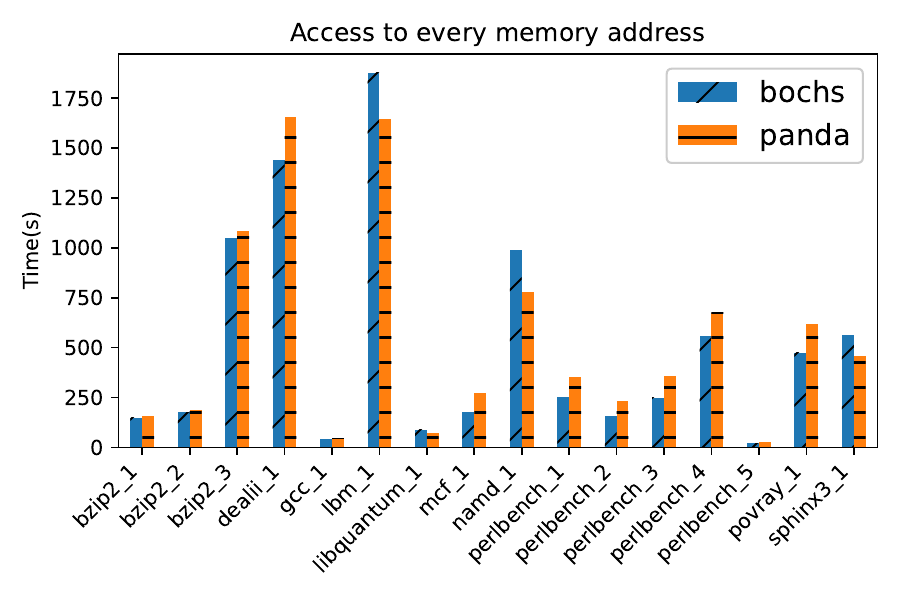}
\caption{Execution time for every memory access instrumentation in benchmark
programs for whole-system tools.}
\label{fig:every_memory_rw_whole}
\end{figure}

Figure~\ref{fig:inscount_whole}, Figure~\ref{fig:address_accesses_whole}, and
Figure~\ref{fig:every_memory_rw_whole} show the measured execution times for
PANDA and Bochs when instrumenting every instruction, accesses to a specific
memory address, and every memory access respectively.

Table~\ref{tbl:single_process} and Table~\ref{tbl:whole_system} present the
average and standard deviation execution times for each experiment in our
evaluation for process-level and whole-system implementations respectively.

\setlength{\tabcolsep}{4pt}

\begin{table*}[htbp]
     \centering
     \tiny
     \caption{Average and standard deviation execution times in seconds for process-level
     technique implementations}  \label{tbl:single_process}
\rotatebox{90}{
     \begin{tabular}{c c r r r r r r r r r r r r r r r r}
     \toprule
      & & \textbf{perlbench\_1} & \textbf{perlbench\_2} & \textbf{perlbench\_3} & \textbf{perlbench\_4} & \textbf{perlbench\_5} & \textbf{bzip2\_1} & \textbf{bzip2\_2} & \textbf{bzip2\_3} & \textbf{gcc} & \textbf{mcf} & \textbf{namd} & \textbf{dealii} & \textbf{povray} & \textbf{libquantum} & \textbf{lbm} & \textbf{sphinx3}\\
   \cmidrule(l){1-2}      \cmidrule(l){3-3} \cmidrule(l){4-4} \cmidrule(l){5-5} \cmidrule(l){6-6} \cmidrule(l){7-7} \cmidrule(l){8-8} \cmidrule(l){9-9} \cmidrule(l){10-10} \cmidrule(l){11-11} \cmidrule(l){12-12} \cmidrule(l){13-13} \cmidrule(l){14-14} \cmidrule(l){15-15} \cmidrule(l){16-16} \cmidrule(l){17-17} \cmidrule(l){18-18}
    \multirow{5}{*}{\textit{exec\_single}} & pin & 5.69$\pm$0.14 & 4.51$\pm$0.16 & 4.64$\pm$0.14 & 13.11$\pm$0.1 & 0.86$\pm$0.02 & 3.15$\pm$0.04 & 3.14$\pm$0.04 & 24.37$\pm$0.16 & 5.16$\pm$0.01 & 7.65$\pm$0.12 & 7.0$\pm$0.06 & 22.62$\pm$0.2 & 7.08$\pm$0.14 & 1.38$\pm$0.02 & 16.05$\pm$0.11 & 5.85$\pm$0.02\\
      & dyninst & 1.85$\pm$0.05 & 1.55$\pm$0.04 & 1.71$\pm$0.04 & 4.6$\pm$0.04 & 0.23$\pm$0.04 & 2.37$\pm$0.05 & 2.34$\pm$0.06 & 18.8$\pm$0.08 & 0.45$\pm$0.06 & 6.98$\pm$0.11 & 6.36$\pm$0.05 & 12.19$\pm$0.06 & 3.04$\pm$0.04 & 0.87$\pm$0.05 & 14.99$\pm$0.06 & 3.96$\pm$0.06\\
      & gdb\_hb & 102.49$\pm$1.02 & 124.21$\pm$0.77 & 110.31$\pm$0.42 & 38.69$\pm$0.33 & 7.73$\pm$0.07 & 4.89$\pm$0.04 & 4.19$\pm$0.03 & 37.06$\pm$0.15 & 3.41$\pm$0.02 & 54.99$\pm$0.3 & 8.77$\pm$0.02 & 57.38$\pm$0.14 & 9.0$\pm$0.05 & 5.72$\pm$0.05 & 29.3$\pm$0.13 & 74.43$\pm$0.43\\
      & gdb\_sb & 99.86$\pm$0.57 & 120.3$\pm$0.89 & 107.32$\pm$0.66 & 38.45$\pm$0.13 & 7.55$\pm$0.05 & 4.94$\pm$0.03 & 4.28$\pm$0.03 & 37.6$\pm$0.15 & 3.4$\pm$0.04 & 54.06$\pm$0.47 & 9.06$\pm$0.04 & 57.43$\pm$0.37 & 8.98$\pm$0.1 & 5.73$\pm$0.06 & 29.17$\pm$0.18 & 73.68$\pm$0.34\\
      & gdb\_pf & 255.73$\pm$0.66 & 254.9$\pm$0.35 & 269.65$\pm$1.98 & 257.33$\pm$2.32 & 269.03$\pm$1.19 & 220.4$\pm$1.38 & 222.92$\pm$1.03 & 220.11$\pm$0.88 & 227.25$\pm$1.1 & 220.29$\pm$1.38 & 271.5$\pm$1.25 & 417.03$\pm$2.16 & 294.43$\pm$0.94 & 252.77$\pm$0.78 & 252.13$\pm$1.14 & 265.91$\pm$1.66\\
      \cmidrule(l){1-2}    \cmidrule(l){3-3} \cmidrule(l){4-4} \cmidrule(l){5-5} \cmidrule(l){6-6} \cmidrule(l){7-7} \cmidrule(l){8-8} \cmidrule(l){9-9} \cmidrule(l){10-10} \cmidrule(l){11-11} \cmidrule(l){12-12} \cmidrule(l){13-13} \cmidrule(l){14-14} \cmidrule(l){15-15} \cmidrule(l){16-16} \cmidrule(l){17-17} \cmidrule(l){18-18}
      \multirow{2}{*}{\textit{exec\_range}} & pin & 5.71$\pm$0.17 & 4.51$\pm$0.04 & 4.8$\pm$0.16 & 13.0$\pm$0.18 & 0.86$\pm$0.03 & 3.11$\pm$0.02 & 3.15$\pm$0.04 & 24.8$\pm$0.26 & 5.33$\pm$0.05 & 18.13$\pm$0.04 & 10.12$\pm$0.07 & 59.26$\pm$0.15 & 6.95$\pm$0.17 & 1.04$\pm$0.02 & 113.93$\pm$0.61 & 5.94$\pm$0.09\\
      & gdb\_pf & 250.52$\pm$1.07 & 249.57$\pm$1.2 & 265.72$\pm$0.98 & 250.86$\pm$2.17 & 263.4$\pm$0.69 & 215.4$\pm$1.39 & 219.16$\pm$0.52 & 215.32$\pm$0.85 & 222.39$\pm$0.78 & 216.49$\pm$0.46 & 266.43$\pm$0.93 & 411.65$\pm$1.1 & 289.37$\pm$1.13 & 248.0$\pm$1.21 & 248.15$\pm$0.73 & 260.08$\pm$1.22\\
     \cmidrule(l){1-2}    \cmidrule(l){3-3} \cmidrule(l){4-4} \cmidrule(l){5-5} \cmidrule(l){6-6} \cmidrule(l){7-7} \cmidrule(l){8-8} \cmidrule(l){9-9} \cmidrule(l){10-10} \cmidrule(l){11-11} \cmidrule(l){12-12} \cmidrule(l){13-13} \cmidrule(l){14-14} \cmidrule(l){15-15} \cmidrule(l){16-16} \cmidrule(l){17-17} \cmidrule(l){18-18}
 \multirow{2}{*}{\textit{exec\_all}}  & pin & 28.92$\pm$0.12 & 15.57$\pm$0.09 & 24.67$\pm$0.13 & 77.94$\pm$0.47 & 2.71$\pm$0.0 & 23.67$\pm$0.07 & 27.93$\pm$0.02 & 167.46$\pm$0.04 & 10.46$\pm$0.03 & 25.92$\pm$0.15 & 73.35$\pm$0.22 & 162.7$\pm$0.4 & 40.92$\pm$0.02 & 14.07$\pm$0.0 & 113.93$\pm$0.25 & 50.7$\pm$0.12\\
      & gdb\_ss & 252.66$\pm$1.92 & 248.18$\pm$1.69 & 249.27$\pm$2.03 & 252.94$\pm$1.18 & 249.46$\pm$1.35 & 197.45$\pm$0.79 & 197.12$\pm$1.47 & 192.86$\pm$1.74 & 287.93$\pm$1.73 & 220.41$\pm$0.83 & 231.83$\pm$1.01 & 304.51$\pm$3.64 & 249.28$\pm$0.83 & 193.99$\pm$0.85 & 187.94$\pm$1.55 & 263.23$\pm$1.52\\
     \cmidrule(l){1-2}    \cmidrule(l){3-3} \cmidrule(l){4-4} \cmidrule(l){5-5} \cmidrule(l){6-6} \cmidrule(l){7-7} \cmidrule(l){8-8} \cmidrule(l){9-9} \cmidrule(l){10-10} \cmidrule(l){11-11} \cmidrule(l){12-12} \cmidrule(l){13-13} \cmidrule(l){14-14} \cmidrule(l){15-15} \cmidrule(l){16-16} \cmidrule(l){17-17} \cmidrule(l){18-18}
     \multirow{3}{*}{\textit{rw\_single}} & pin & 39.72$\pm$0.22 & 29.33$\pm$0.05 & 36.66$\pm$1.32 & 87.36$\pm$0.04 & 4.03$\pm$0.01 & 23.84$\pm$0.0 & 26.29$\pm$0.02 & 167.4$\pm$0.46 & 62.28$\pm$0.12 & 24.56$\pm$0.04 & 54.56$\pm$0.01 & 165.27$\pm$0.08 & 57.12$\pm$0.16 & 11.22$\pm$0.01 & 89.77$\pm$0.01 & 81.29$\pm$0.01\\
     & gdb\_hb & 94.76$\pm$0.68 & 26.53$\pm$0.16 & 1.89$\pm$0.0 & 150.84$\pm$0.53 & 0.69$\pm$0.0 & 28.28$\pm$0.11 & 15.09$\pm$0.13 & 224.19$\pm$0.95 & 4.52$\pm$0.05 & 42.41$\pm$0.25 & 8.86$\pm$0.03 & 59.31$\pm$0.36 & 4.4$\pm$0.03 & 4.58$\pm$0.04 & 16.58$\pm$0.63 & 59.48$\pm$0.27\\
     & gdb\_pf & 268.85$\pm$1.03 & 268.55$\pm$1.14 & 284.35$\pm$0.89 & 266.91$\pm$4.38 & 282.37$\pm$1.13 & 222.73$\pm$12.67 & 218.72$\pm$1.36 & 215.54$\pm$2.58 & 245.02$\pm$1.59 & 220.2$\pm$0.85 & 275.6$\pm$1.41 & 373.42$\pm$1.01 & 313.81$\pm$1.64 & 251.35$\pm$1.28 & 19.44$\pm$0.43 & 269.07$\pm$1.35\\
     \cmidrule(l){1-2}    \cmidrule(l){3-3} \cmidrule(l){4-4} \cmidrule(l){5-5} \cmidrule(l){6-6} \cmidrule(l){7-7} \cmidrule(l){8-8} \cmidrule(l){9-9} \cmidrule(l){10-10} \cmidrule(l){11-11} \cmidrule(l){12-12} \cmidrule(l){13-13} \cmidrule(l){14-14} \cmidrule(l){15-15} \cmidrule(l){16-16} \cmidrule(l){17-17} \cmidrule(l){18-18}

    \multirow{2}{*}{\textit{rw\_range}} & pin & 42.43$\pm$0.03 & 31.68$\pm$0.04 & 38.86$\pm$0.07 & 97.29$\pm$0.02 & 4.34$\pm$0.01 & 27.15$\pm$0.0 & 30.32$\pm$0.01 & 191.9$\pm$0.2 & 62.8$\pm$0.02 & 28.92$\pm$0.03 & 62.53$\pm$0.01 & 192.46$\pm$0.36 & 64.15$\pm$0.28 & 13.05$\pm$0.0 & 102.7$\pm$0.02 & 87.56$\pm$0.12\\
     & gdb\_pf & 262.64$\pm$0.26 & 262.83$\pm$0.83 & 277.98$\pm$1.62 & 262.6$\pm$0.66 & 275.79$\pm$3.2 & 211.53$\pm$3.09 & 211.47$\pm$1.83 & 203.41$\pm$6.77 & 239.24$\pm$1.12 & 214.24$\pm$1.46 & 262.57$\pm$1.84 & 374.95$\pm$1.13 & 283.52$\pm$2.17 & 246.96$\pm$0.69 & 19.47$\pm$0.48 & 263.37$\pm$0.79\\
     \cmidrule(l){1-2}    \cmidrule(l){3-3} \cmidrule(l){4-4} \cmidrule(l){5-5} \cmidrule(l){6-6} \cmidrule(l){7-7} \cmidrule(l){8-8} \cmidrule(l){9-9} \cmidrule(l){10-10} \cmidrule(l){11-11} \cmidrule(l){12-12} \cmidrule(l){13-13} \cmidrule(l){14-14} \cmidrule(l){15-15} \cmidrule(l){16-16} \cmidrule(l){17-17} \cmidrule(l){18-18}
      \textit{rw\_all} & pin & 15.88$\pm$0.04 & 8.13$\pm$0.03 & 13.16$\pm$0.03 & 118.05$\pm$0.52 & 1.82$\pm$0.02 & 18.32$\pm$0.01 & 17.73$\pm$0.01 & 128.25$\pm$0.17 & 59.54$\pm$0.02 & 12.19$\pm$0.1 & 20.59$\pm$0.09 & 67.29$\pm$0.03 & 31.67$\pm$0.01 & 5.16$\pm$0.0 & 36.98$\pm$0.09 & 63.51$\pm$0.35\\
     \cmidrule(l){1-2}    \cmidrule(l){3-3} \cmidrule(l){4-4} \cmidrule(l){5-5} \cmidrule(l){6-6} \cmidrule(l){7-7} \cmidrule(l){8-8} \cmidrule(l){9-9} \cmidrule(l){10-10} \cmidrule(l){11-11} \cmidrule(l){12-12} \cmidrule(l){13-13} \cmidrule(l){14-14} \cmidrule(l){15-15} \cmidrule(l){16-16} \cmidrule(l){17-17} \cmidrule(l){18-18}
     \multirow{3}{*}{\textit{none}} & pin & 5.33$\pm$0.02 & 4.15$\pm$0.02 & 4.43$\pm$0.05 & 12.19$\pm$0.02 & 0.79$\pm$0.0 & 3.02$\pm$0.01 & 2.96$\pm$0.0 & 23.04$\pm$0.04 & 4.93$\pm$0.02 & 7.22$\pm$0.02 & 6.75$\pm$0.02 & 21.56$\pm$0.03 & 6.73$\pm$0.02 & 0.97$\pm$0.01 & 15.23$\pm$0.09 & 5.11$\pm$0.0\\
      & dyninst & 1.79$\pm$0.01 & 1.48$\pm$0.01 & 1.65$\pm$0.0 & 4.58$\pm$0.01 & 0.16$\pm$0.01 & 2.3$\pm$0.01 & 2.28$\pm$0.02 & 18.76$\pm$0.12 & 0.41$\pm$0.01 & 6.95$\pm$0.16 & 6.34$\pm$0.01 & 12.05$\pm$0.04 & 2.95$\pm$0.02 & 0.78$\pm$0.01 & 15.02$\pm$0.15 & 3.82$\pm$0.01\\
      & gdb & 1.8$\pm$0.0 & 1.52$\pm$0.02 & 1.66$\pm$0.01 & 4.57$\pm$0.0 & 0.16$\pm$0.01 & 2.32$\pm$0.01 & 2.3$\pm$0.02 & 18.76$\pm$0.09 & 0.4$\pm$0.0 & 6.84$\pm$0.04 & 6.34$\pm$0.04 & 12.08$\pm$0.03 & 2.96$\pm$0.02 & 0.77$\pm$0.01 & 14.87$\pm$0.05 & 3.85$\pm$0.02\\
     \bottomrule
     \end{tabular}
     }
     \end{table*}

\setlength{\tabcolsep}{3pt}

\begin{table*}[htbp]
  \centering
  \tiny
  \caption{Average and standard deviation execution times in seconds for whole-system
     technique implementations}  \label{tbl:whole_system}
\rotatebox{90}{
  \begin{tabular}{c c r r r r r r r r r r r r r r r r }
  \toprule
&   & \textbf{perlbench\_1} & \textbf{perlbench\_2} & \textbf{perlbench\_3} & \textbf{perlbench\_4} & \textbf{perlbench\_5} & \textbf{bzip2\_1} & \textbf{bzip2\_2} & \textbf{bzip2\_3} & \textbf{gcc} & \textbf{mcf} & \textbf{namd} & \textbf{dealii} & \textbf{povray} & \textbf{libquantum} & \textbf{lbm} & \textbf{sphinx3}\\
    \cmidrule(l){1-2}      \cmidrule(l){3-3} \cmidrule(l){4-4} \cmidrule(l){5-5} \cmidrule(l){6-6} \cmidrule(l){7-7} \cmidrule(l){8-8} \cmidrule(l){9-9} \cmidrule(l){10-10} \cmidrule(l){11-11} \cmidrule(l){12-12} \cmidrule(l){13-13} \cmidrule(l){14-14} \cmidrule(l){15-15} \cmidrule(l){16-16} \cmidrule(l){17-17} \cmidrule(l){18-18}
    \multirow{7}{*}{\textit{exec\_single}} & bochs & 233.54$\pm$5.21 & 147.7$\pm$0.47 & 215.29$\pm$1.43 & 517.84$\pm$10.59 & 18.81$\pm$0.11 & 134.72$\pm$0.45 & 160.86$\pm$0.29 & 959.99$\pm$8.38 & 38.05$\pm$0.31 & 169.9$\pm$1.36 & 962.24$\pm$2.91 & 1346.45$\pm$7.27 & 442.25$\pm$1.56 & 81.57$\pm$0.61 & 1770.39$\pm$3.19 & 541.87$\pm$1.29\\
    & panda & 97.51$\pm$0.69 & 62.01$\pm$0.64 & 81.87$\pm$0.52 & 189.37$\pm$0.88 & 7.21$\pm$0.06 & 39.37$\pm$0.48 & 46.31$\pm$0.4 & 308.86$\pm$0.93 & 12.7$\pm$0.16 & 141.11$\pm$1.01 & 473.89$\pm$1.15 & 576.05$\pm$1.35 & 230.75$\pm$0.91 & 15.05$\pm$0.22 & 1086.19$\pm$154.73 & 252.08$\pm$2.13\\
    & drakvuf\_sb & 43.33$\pm$0.96 & 50.64$\pm$0.44 & 45.23$\pm$0.81 & 19.17$\pm$0.36 & 3.77$\pm$0.2 & 3.72$\pm$0.08 & 3.4$\pm$0.08 & 26.37$\pm$0.24 & 2.15$\pm$0.11 & 28.58$\pm$0.67 & 7.61$\pm$0.05 & 30.68$\pm$0.35 & 6.7$\pm$0.17 & 3.1$\pm$0.06 & 21.04$\pm$0.25 & 32.92$\pm$0.79\\
    & drakvuf\_pf & 11.88$\pm$0.25 & 11.54$\pm$0.22 & 11.53$\pm$0.39 & 11.6$\pm$0.4 & 11.12$\pm$0.3 & 11.71$\pm$0.28 & 11.76$\pm$0.47 & 11.7$\pm$0.39 & 11.3$\pm$0.44 & 11.54$\pm$0.43 & 12.18$\pm$0.23 & 11.46$\pm$0.22 & 12.19$\pm$0.3 & 12.21$\pm$0.27 & 15.75$\pm$0.4 & 12.01$\pm$0.23\\
    \cmidrule(l){1-2}      \cmidrule(l){3-3} \cmidrule(l){4-4} \cmidrule(l){5-5} \cmidrule(l){6-6} \cmidrule(l){7-7} \cmidrule(l){8-8} \cmidrule(l){9-9} \cmidrule(l){10-10} \cmidrule(l){11-11} \cmidrule(l){12-12} \cmidrule(l){13-13} \cmidrule(l){14-14} \cmidrule(l){15-15} \cmidrule(l){16-16} \cmidrule(l){17-17} \cmidrule(l){18-18}
   \multirow{5}{*}{\textit{exec\_range}}  & bochs & 247.41$\pm$2.38 & 158.65$\pm$0.4 & 224.04$\pm$0.66 & 545.88$\pm$40.03 & 19.82$\pm$0.09 & 141.93$\pm$0.56 & 168.75$\pm$0.29 & 999.32$\pm$1.54 & 38.84$\pm$0.14 & 171.69$\pm$0.26 & 988.18$\pm$0.77 & 1289.17$\pm$214.01 & 462.76$\pm$0.6 & 85.4$\pm$0.2 & 1856.45$\pm$1.66 & 570.91$\pm$1.14\\
     & panda & 98.22$\pm$0.94 & 66.69$\pm$0.93 & 83.73$\pm$0.65 & 193.62$\pm$1.16 & 7.31$\pm$0.08 & 38.54$\pm$0.27 & 45.42$\pm$0.2 & 305.21$\pm$1.21 & 13.77$\pm$0.15 & 211.56$\pm$1.15 & 493.02$\pm$2.23 & 816.7$\pm$6.92 & 222.99$\pm$0.24 & 14.63$\pm$0.03 & 2399.87$\pm$168.04 & 248.8$\pm$0.21\\
     & drakvuf\_pf & 11.87$\pm$0.19 & 11.52$\pm$0.26 & 11.55$\pm$0.15 & 12.07$\pm$0.42 & 11.51$\pm$0.27 & 11.7$\pm$0.3 & 11.76$\pm$0.47 & 11.7$\pm$0.39 & 11.52$\pm$0.25 & 11.41$\pm$0.17 & 12.83$\pm$0.24 & 11.47$\pm$0.21 & 12.18$\pm$0.27 & 12.07$\pm$0.26 & 15.68$\pm$0.4 & 12.14$\pm$0.16\\
     \cmidrule(l){1-2}      \cmidrule(l){3-3} \cmidrule(l){4-4} \cmidrule(l){5-5} \cmidrule(l){6-6} \cmidrule(l){7-7} \cmidrule(l){8-8} \cmidrule(l){9-9} \cmidrule(l){10-10} \cmidrule(l){11-11} \cmidrule(l){12-12} \cmidrule(l){13-13} \cmidrule(l){14-14} \cmidrule(l){15-15} \cmidrule(l){16-16} \cmidrule(l){17-17} \cmidrule(l){18-18}  
    \multirow{3}{*}{\textit{exec\_all}} & bochs & 274.23$\pm$90.12 & 148.81$\pm$0.31 & 216.38$\pm$0.48 & 522.18$\pm$8.8 & 18.96$\pm$0.08 & 140.6$\pm$0.22 & 166.22$\pm$0.45 & 985.46$\pm$6.92 & 37.21$\pm$0.09 & 167.29$\pm$0.77 & 961.83$\pm$4.06 & 1381.23$\pm$3.33 & 451.63$\pm$1.34 & 84.93$\pm$0.3 & 1817.49$\pm$3.03 & 563.86$\pm$0.86\\
    & panda & 359.66$\pm$3.95 & 234.80$\pm$0.6 & 353.85$\pm$0.64 & 682.9$\pm$8.7 & 29.21$\pm$0.21 & 148.37$\pm$0.81 & 165.62$\pm$0.29 & 1058.66$\pm$3.45 & 45.44$\pm$3.93 & 239.78$\pm$0.52 & 933.63$\pm$6.33 & 1350.81$\pm$12.15 & 759.08$\pm$252.91 & 70.13$\pm$0.29 & 2318.74$\pm$29.7 & 498.34$\pm$6.47\\
    \cmidrule(l){1-2}      \cmidrule(l){3-3} \cmidrule(l){4-4} \cmidrule(l){5-5} \cmidrule(l){6-6} \cmidrule(l){7-7} \cmidrule(l){8-8} \cmidrule(l){9-9} \cmidrule(l){10-10} \cmidrule(l){11-11} \cmidrule(l){12-12} \cmidrule(l){13-13} \cmidrule(l){14-14} \cmidrule(l){15-15} \cmidrule(l){16-16} \cmidrule(l){17-17} \cmidrule(l){18-18}
     \multirow{3}{*}{\textit{rw\_single}} & bochs & 241.88$\pm$1.39 & 152.37$\pm$0.33 & 224.56$\pm$0.59 & 530.32$\pm$0.77 & 19.39$\pm$0.32 & 139.81$\pm$0.37 & 167.54$\pm$0.47 & 982.88$\pm$1.89 & 39.8$\pm$0.22 & 170.8$\pm$0.66 & 960.09$\pm$1.29 & 1358.85$\pm$1.83 & 455.34$\pm$1.17 & 82.13$\pm$0.28 & 1847.86$\pm$1.87 & 546.95$\pm$1.03\\
  & panda & 359.13$\pm$1.44 & 240.61$\pm$0.68 & 366.06$\pm$3.25 & 676.34$\pm$4.43 & 28.88$\pm$0.22 & 161.53$\pm$1.47 & 187.1$\pm$1.63 & 1154.18$\pm$15.47 & 40.45$\pm$0.16 & 275.58$\pm$0.64 & 790.13$\pm$4.19 & 1557.24$\pm$7.28 & 620.89$\pm$0.72 & 72.31$\pm$0.21 & 1591.37$\pm$26.92 & 524.35$\pm$150.65\\
   \cmidrule(l){1-2}      \cmidrule(l){3-3} \cmidrule(l){4-4} \cmidrule(l){5-5} \cmidrule(l){6-6} \cmidrule(l){7-7} \cmidrule(l){8-8} \cmidrule(l){9-9} \cmidrule(l){10-10} \cmidrule(l){11-11} \cmidrule(l){12-12} \cmidrule(l){13-13} \cmidrule(l){14-14} \cmidrule(l){15-15} \cmidrule(l){16-16} \cmidrule(l){17-17} \cmidrule(l){18-18}
   \multirow{5}{*}{\textit{rw\_range}} & bochs & 293.06$\pm$81.98 & 162.74$\pm$0.47 & 231.22$\pm$0.67 & 538.98$\pm$1.15 & 19.72$\pm$0.09 & 145.45$\pm$1.18 & 174.2$\pm$0.89 & 1028.29$\pm$5.49 & 40.59$\pm$0.31 & 175.62$\pm$1.07 & 980.12$\pm$2.47 & 1417.88$\pm$7.79 & 469.97$\pm$0.75 & 84.22$\pm$0.2 & 1870.98$\pm$2.9 & 552.04$\pm$0.97\\
   & panda & 359.5$\pm$1.07 & 237.98$\pm$1.31 & 360.21$\pm$0.46 & 676.15$\pm$0.78 & 28.59$\pm$0.06 & 160.49$\pm$0.33 & 189.94$\pm$7.3 & 1150.98$\pm$3.32 & 40.13$\pm$0.2 & 274.51$\pm$0.84 & 829.76$\pm$116.76 & 1667.16$\pm$272.88 & 625.65$\pm$8.26 & 73.33$\pm$1.22 & 1665.82$\pm$27.52 & 457.44$\pm$1.19\\
  & drakvuf\_pf & 11.68$\pm$0.18 & 11.41$\pm$0.2 & 11.56$\pm$0.15 & 11.6$\pm$0.23 & 11.65$\pm$0.3 & 11.7$\pm$0.32 & 11.53$\pm$0.24 & 11.49$\pm$0.24 & 11.59$\pm$0.2 & 11.63$\pm$0.28 & 12.42$\pm$0.19 & 12.56$\pm$0.31 & 12.18$\pm$0.27 & 12.18$\pm$0.09 & 15.58$\pm$0.04 & 12.67$\pm$0.13\\
  \cmidrule(l){1-2}      \cmidrule(l){3-3} \cmidrule(l){4-4} \cmidrule(l){5-5} \cmidrule(l){6-6} \cmidrule(l){7-7} \cmidrule(l){8-8} \cmidrule(l){9-9} \cmidrule(l){10-10} \cmidrule(l){11-11} \cmidrule(l){12-12} \cmidrule(l){13-13} \cmidrule(l){14-14} \cmidrule(l){15-15} \cmidrule(l){16-16} \cmidrule(l){17-17} \cmidrule(l){18-18}
     \multirow{3}{*}{\textit{rw\_all}} & bochs & 250.55$\pm$1.05 & 156.36$\pm$0.76 & 248.28$\pm$5.61 & 558.71$\pm$4.85 & 20.06$\pm$0.18 & 148.66$\pm$0.26 & 176.86$\pm$0.33 & 1047.44$\pm$5.81 & 40.63$\pm$0.09 & 178.44$\pm$0.5 & 988.54$\pm$0.88 & 1441.05$\pm$28.26 & 474.37$\pm$3.88 & 87.7$\pm$0.37 & 1877.04$\pm$4.07 & 564.31$\pm$1.13\\
     & panda & 351.4$\pm$1.08 & 233.37$\pm$0.6 & 357.57$\pm$3.47 & 673.52$\pm$8.34 & 28.85$\pm$0.21 & 159.09$\pm$0.5 & 185.54$\pm$0.32 & 1086.66$\pm$25.52 & 42.33$\pm$7.94 & 272.55$\pm$0.56 & 778.57$\pm$1.7 & 1656.88$\pm$295.07 & 618.32$\pm$7.8 & 70.92$\pm$0.15 & 1643.76$\pm$11.03 & 459.34$\pm$8.57\\
     \cmidrule(l){1-2}      \cmidrule(l){3-3} \cmidrule(l){4-4} \cmidrule(l){5-5} \cmidrule(l){6-6} \cmidrule(l){7-7} \cmidrule(l){8-8} \cmidrule(l){9-9} \cmidrule(l){10-10} \cmidrule(l){11-11} \cmidrule(l){12-12} \cmidrule(l){13-13} \cmidrule(l){14-14} \cmidrule(l){15-15} \cmidrule(l){16-16} \cmidrule(l){17-17} \cmidrule(l){18-18}
  \multirow{5}{*}{\textit{none}}    & bochs & 151.03$\pm$1.44 & 112.08$\pm$0.58 & 149.17$\pm$0.79 & 308.13$\pm$0.79 & 13.23$\pm$0.15 & 80.23$\pm$0.52 & 97.4$\pm$3.72 & 558.03$\pm$3.59 & 27.53$\pm$0.37 & 105.06$\pm$0.49 & 742.31$\pm$4.8 & 885.86$\pm$4.55 & 320.21$\pm$0.86 & 46.96$\pm$0.42 & 1432.52$\pm$2.2 & 411.81$\pm$9.91\\
     & panda & 97.26$\pm$1.1 & 62.73$\pm$0.65 & 82.08$\pm$0.91 & 190.05$\pm$1.05 & 7.32$\pm$0.06 & 38.56$\pm$0.21 & 45.95$\pm$0.61 & 307.39$\pm$1.38 & 12.86$\pm$0.18 & 142.34$\pm$1.7 & 460.18$\pm$0.56 & 559.14$\pm$2.44 & 226.39$\pm$0.34 & 14.72$\pm$0.09 & 1003.73$\pm$0.6 & 249.56$\pm$1.6\\
     & drakvuf & 2.2$\pm$0.02 & 1.79$\pm$0.01 & 1.91$\pm$0.02 & 5.05$\pm$0.09 & 0.41$\pm$0.01 & 2.47$\pm$0.03 & 2.44$\pm$0.03 & 19.11$\pm$1.95 & 0.67$\pm$0.01 & 7.42$\pm$0.06 & 6.53$\pm$0.03 & 12.48$\pm$0.05 & 3.79$\pm$0.04 & 0.96$\pm$0.01 & 14.96$\pm$0.28 & 4.16$\pm$0.06\\
\bottomrule
  \end{tabular}
}
\end{table*}

\end{document}